\documentclass[longauth]{aa}

\usepackage{graphicx}
\usepackage{natbib}
\usepackage{scalerel}

\usepackage[table]{xcolor}
\usepackage{makecell}

\usepackage[varg]{txfonts}
\usepackage[pdfencoding=auto,psdextra]{hyperref}
\usepackage{orcidlink}
\usepackage{ulem}
\usepackage{xcolor}
\usepackage{cancel}
\newcommand*{\omm}{\ensuremath{\omega_{\rm m}}}
\newcommand*{\omb}{\ensuremath{\omega_{\rm b}}}
\newcommand{\thbao}{\ensuremath{\theta_\mathrm{BAO}}\xspace}
\newcommand*{\thmin}{\ensuremath{\theta_{\rm min}}}
\newcommand*{\thmax}{\ensuremath{\theta_{\rm max}}}
\usepackage{enumitem}
\setlist[itemize]{topsep=0pt,partopsep=0pt,parsep=0pt,itemsep=0pt}
\defcitealias{Abbott_2024_DESY6}{DES Y6}

\hypersetup{
    colorlinks=true,
    linkcolor=blue,
    filecolor=magenta,      
    urlcolor=blue,
    citecolor=blue
}
\usepackage{lastpage}
\makeatletter

\AtBeginDocument{\renewcommand*\aa@pageof{, page \thepage{} of \hyperref[LastPage]{\pageref*{LastPage}}}} 

\idline{\aap, 713, A281 (2026)}
\yearCop{2026}
\renewcommand*\aa@copyrightname{\copyright~The Authors}  

\newcommand*\AA@href[3]{\begingroup\hypersetup{urlcolor=#1}\href{#2}{#3}\endgroup}

\newcommand*\AAdoi[1]{\doi{#1}\renewcommand*\aa@doi{\AA@href{blue}{https://doi.org/#1}{https://doi.org/#1}}}

\AAdoi{10.1051/0004-6361/202554545}

\let\AA@classlogo\AALogo
\renewcommand*\AALogo{\AA@href{blue}{https://www.aanda.org}{\AA@classlogo}}

\fancypagestyle{otherpage}{%
  \fancyhf{}%
  \fancyhead[C]{\aa@headfont \aa@headings}%
  \fancyfoot[RO]{\aa@footfont \aa@numarticle\aa@pageof}%
  \fancyfoot[LE]{\aa@footfont \aa@numarticle\aa@pageof}%
  \renewcommand*{\headrulewidth}{\z@}%
  \renewcommand*{\footrulewidth}{\z@}%
}

\newcommand*\AA@openaccessnotice{Open Access article, \href{https://www.edpsciences.org/}{published by EDP Sciences}, under the terms of the Creative Commons Attribution License (\href{https://creativecommons.org/licenses/by/4.0}{https://creativecommons.org/licenses/by/4.0}),\\which permits unrestricted use, distribution, and reproduction in any medium, provided the original work is properly cited.\\This article is published in open access under the \href{https://www.aanda.org/subscribe-to-open-faqs}{Subscribe to Open model}. \href{mailto:subscribers@edpsciences.org}{Subscribe to \aap} to support open access publication.}
\fancypagestyle{firstpage}{%
  \fancyhf{}%
  \renewcommand*{\headrulewidth}{\z@}%
  \renewcommand*{\footrulewidth}{\z@}%
  \fancyfoot[C]{%
    \parbox[t]{\headwidth}{%
      \noindent\hfill{\aa@footfont \aa@numarticle\aa@pageof}\par
      \vspace{3pt}%
      {\scriptsize\centering\AA@openaccessnotice\par}}}%
}
    
\makeatother
\AANum{A281}

\usepackage[utf8]{inputenc}

\usepackage[switch, modulo]{lineno}

\usepackage{euclid}
\renewcommand*{\NDD}{\ensuremath{N_\mathrm{DD}}}
\renewcommand*{\NRR}{\ensuremath{N_\mathrm{RR}}}
\renewcommand*{\NDR}{\ensuremath{N_\mathrm{DR}}}

\begin{document}

\title{\Euclid preparation}
\subtitle{CIX. Baryon acoustic oscillation analysis of photometric galaxy clustering in configuration space}

\renewcommand{\orcidlink}[1]{\href{https://orcid.org/#1}{\raisebox{0.03ex}{\includegraphics[height=1.6ex]{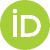}}}}

\newcommand*{\corrauth}[1]{\thanks{Corresponding author: {\ttfamily #1}}}

\makeatletter

\newdimen\AA@fnmarkwd
\DeclareRobustCommand\thanks[1]{
  \footnotemark
  \begingroup
    \reset@font\footnotesize
    \setbox\@tempboxa\hbox{\@makefnmark}%
    \ifdim\wd\@tempboxa>\AA@fnmarkwd \global\AA@fnmarkwd=\wd\@tempboxa\fi
  \endgroup
  \protected@xdef\@thanks{\@thanks
    \protect\AA@thankstext{\the\c@footnote}{\Hy@footnote@currentHref}{#1}}}
    
\newcommand\AA@thankstext[3]{%
  \begingroup
    \def\@makefntext##1{%
      \noindent
      \hb@xt@\bibindent{\hss\hb@xt@\AA@fnmarkwd{\@makefnmark\hss}\enspace}##1}%
    \footnotetext[#1]{%
      \Hy@raisedlink{\hyper@@anchor{#2}{\relax}}
      \ignorespaces #3}%
  \endgroup}
  
\DeclareRobustCommand\AAfnstore[1]{\expandafter\xdef\csname AA@fnnum@#1\endcsname{\the\c@footnote}\expandafter\xdef\csname AA@fnhref@#1\endcsname{\Hy@footnote@currentHref}}

\DeclareRobustCommand\AAfnmark[1]{%
  \@ifundefined{AA@fnhref@#1}%
    {\PackageWarning{AAfn}{Footnote `#1' unknown: \string\AAfnstore\space must be used (after a \string\thanks) before \string\AAfnmark}%
     \textsuperscript{?}}%
    {\begingroup
       \c@footnote=\csname AA@fnnum@#1\endcsname\relax
       \protected@edef\@thefnmark{\thefootnote}%
       \edef\AA@fntarget{\csname AA@fnhref@#1\endcsname}%
       \hyper@linkstart{link}{\AA@fntarget}%
       \@makefnmark
       \hyper@linkend
     \endgroup}}

\renewcommand*\fnmsep{\unskip\hbox{\@textsuperscript{\normalfont,}\kern-0.3em}}
\makeatother

\let\oldinst\inst
\renewcommand{\inst}[1]{{\hypersetup{linkcolor=black}\oldinst{#1}}}

\author{
Euclid Collaboration: V.~Duret\inst{\ref{aff1}}\fnmsep\corrauth{vincent.duret@etu.univ-amu.fr}\orcidlink{0009-0009-0383-4960}
\and S.~Escoffier\inst{\ref{aff1}}\orcidlink{0000-0002-2847-7498}
\and W.~Gillard\inst{\ref{aff1}}\orcidlink{0000-0003-4744-9748}
\and I.~Tutusaus\inst{\ref{aff2}}\orcidlink{0000-0002-3199-0399}
\and S.~Camera\inst{\ref{aff3},\!\ref{aff4},\!\ref{aff5}}\orcidlink{0000-0003-3399-3574}
\and N.~Tessore\inst{\ref{aff6}}\orcidlink{0000-0002-9696-7931}
\and F.~J.~Castander\inst{\ref{aff7},\!\ref{aff8}}\orcidlink{0000-0001-7316-4573}
\and N.~Aghanim\inst{\ref{aff9}}\orcidlink{0000-0002-6688-8992}
\and A.~Amara\inst{\ref{aff10}}
\and L.~Amendola\inst{\ref{aff11}}\orcidlink{0000-0002-0835-233X}
\and S.~Andreon\inst{\ref{aff12}}\orcidlink{0000-0002-2041-8784}
\and N.~Auricchio\inst{\ref{aff13}}\orcidlink{0000-0003-4444-8651}
\and C.~Baccigalupi\inst{\ref{aff14},\!\ref{aff15},\!\ref{aff16},\!\ref{aff17}}\orcidlink{0000-0002-8211-1630}
\and M.~Baldi\inst{\ref{aff18},\!\ref{aff13},\!\ref{aff19}}\orcidlink{0000-0003-4145-1943}
\and S.~Bardelli\inst{\ref{aff13}}\orcidlink{0000-0002-8900-0298}
\and P.~Battaglia\inst{\ref{aff13}}\orcidlink{0000-0002-7337-5909}
\and A.~Biviano\inst{\ref{aff15},\!\ref{aff14}}\orcidlink{0000-0002-0857-0732}
\and D.~Bonino\inst{\ref{aff5}}\orcidlink{0000-0002-3336-9977}
\and E.~Branchini\inst{\ref{aff20},\!\ref{aff21},\!\ref{aff12}}\orcidlink{0000-0002-0808-6908}
\and M.~Brescia\inst{\ref{aff22},\!\ref{aff23},\!\ref{aff24}}\orcidlink{0000-0001-9506-5680}
\and J.~Brinchmann\inst{\ref{aff25},\!\ref{aff26}}\orcidlink{0000-0003-4359-8797}
\and A.~Caillat\inst{\ref{aff27}}
\and G.~Ca\~nas-Herrera\inst{\ref{aff28},\!\ref{aff29}}\orcidlink{0000-0003-2796-2149}
\and V.~Capobianco\inst{\ref{aff5}}\orcidlink{0000-0002-3309-7692}
\and C.~Carbone\inst{\ref{aff30}}\orcidlink{0000-0003-0125-3563}
\and V.~F.~Cardone\inst{\ref{aff31},\!\ref{aff32}}
\and J.~Carretero\inst{\ref{aff33},\!\ref{aff34}}\orcidlink{0000-0002-3130-0204}
\and S.~Casas\inst{\ref{aff35},\!\ref{aff36}}\orcidlink{0000-0002-4751-5138}
\and M.~Castellano\inst{\ref{aff31}}\orcidlink{0000-0001-9875-8263}
\and G.~Castignani\inst{\ref{aff13}}\orcidlink{0000-0001-6831-0687}
\and S.~Cavuoti\inst{\ref{aff23},\!\ref{aff24}}\orcidlink{0000-0002-3787-4196}
\and K.~C.~Chambers\inst{\ref{aff37}}\orcidlink{0000-0001-6965-7789}
\and A.~Cimatti\inst{\ref{aff38}}
\and C.~Colodro-Conde\inst{\ref{aff39}}
\and G.~Congedo\inst{\ref{aff40}}\orcidlink{0000-0003-2508-0046}
\and C.~J.~Conselice\inst{\ref{aff41}}\orcidlink{0000-0003-1949-7638}
\and L.~Conversi\inst{\ref{aff42},\!\ref{aff43}}\orcidlink{0000-0002-6710-8476}
\and Y.~Copin\inst{\ref{aff44}}\orcidlink{0000-0002-5317-7518}
\and F.~Courbin\inst{\ref{aff45},\!\ref{aff46}}\orcidlink{0000-0003-0758-6510}
\and H.~M.~Courtois\inst{\ref{aff47}}\orcidlink{0000-0003-0509-1776}
\and M.~Cropper\inst{\ref{aff48}}\orcidlink{0000-0003-4571-9468}
\and A.~Da~Silva\inst{\ref{aff49},\!\ref{aff50}}\orcidlink{0000-0002-6385-1609}
\and H.~Degaudenzi\inst{\ref{aff51}}\orcidlink{0000-0002-5887-6799}
\and S.~de~la~Torre\inst{\ref{aff27}}
\and G.~De~Lucia\inst{\ref{aff15}}\orcidlink{0000-0002-6220-9104}
\and A.~M.~Di~Giorgio\inst{\ref{aff52}}\orcidlink{0000-0002-4767-2360}
\and H.~Dole\inst{\ref{aff9}}\orcidlink{0000-0002-9767-3839}
\and F.~Dubath\inst{\ref{aff51}}\orcidlink{0000-0002-6533-2810}
\and X.~Dupac\inst{\ref{aff43}}
\and S.~Dusini\inst{\ref{aff53}}\orcidlink{0000-0002-1128-0664}
\and A.~Ealet\inst{\ref{aff44}}\orcidlink{0000-0003-3070-014X}
\and M.~Farina\inst{\ref{aff52}}\orcidlink{0000-0002-3089-7846}
\and R.~Farinelli\inst{\ref{aff13}}
\and S.~Farrens\inst{\ref{aff54}}\orcidlink{0000-0002-9594-9387}
\and F.~Faustini\inst{\ref{aff55},\!\ref{aff31}}\orcidlink{0000-0001-6274-5145}
\and S.~Ferriol\inst{\ref{aff44}}
\and F.~Finelli\inst{\ref{aff13},\!\ref{aff56}}\orcidlink{0000-0002-6694-3269}
\and S.~Fotopoulou\inst{\ref{aff57}}\orcidlink{0000-0002-9686-254X}
\and N.~Fourmanoit\inst{\ref{aff1}}\orcidlink{0009-0005-6816-6925}
\and M.~Frailis\inst{\ref{aff15}}\orcidlink{0000-0002-7400-2135}
\and E.~Franceschi\inst{\ref{aff13}}\orcidlink{0000-0002-0585-6591}
\and M.~Fumana\inst{\ref{aff30}}\orcidlink{0000-0001-6787-5950}
\and S.~Galeotta\inst{\ref{aff15}}\orcidlink{0000-0002-3748-5115}
\and B.~Gillis\inst{\ref{aff40}}\orcidlink{0000-0002-4478-1270}
\and C.~Giocoli\inst{\ref{aff13},\!\ref{aff19}}\orcidlink{0000-0002-9590-7961}
\and J.~Gracia-Carpio\inst{\ref{aff58}}
\and A.~Grazian\inst{\ref{aff59}}\orcidlink{0000-0002-5688-0663}
\and F.~Grupp\inst{\ref{aff58},\!\ref{aff60}}
\and S.~V.~H.~Haugan\inst{\ref{aff61}}\orcidlink{0000-0001-9648-7260}
\and W.~Holmes\inst{\ref{aff62}}
\and F.~Hormuth\inst{\ref{aff63}}
\and A.~Hornstrup\inst{\ref{aff64},\!\ref{aff65}}\orcidlink{0000-0002-3363-0936}
\and P.~Hudelot\inst{\ref{aff66}}
\and K.~Jahnke\inst{\ref{aff67}}\orcidlink{0000-0003-3804-2137}
\and M.~Jhabvala\inst{\ref{aff68}}
\and B.~Joachimi\inst{\ref{aff6}}\orcidlink{0000-0001-7494-1303}
\and E.~Keih\"anen\inst{\ref{aff69}}\orcidlink{0000-0003-1804-7715}
\and S.~Kermiche\inst{\ref{aff1}}\orcidlink{0000-0002-0302-5735}
\and A.~Kiessling\inst{\ref{aff62}}\orcidlink{0000-0002-2590-1273}
\and M.~Kilbinger\inst{\ref{aff54}}\orcidlink{0000-0001-9513-7138}
\and B.~Kubik\inst{\ref{aff44}}\orcidlink{0009-0006-5823-4880}
\and M.~Kunz\inst{\ref{aff70}}\orcidlink{0000-0002-3052-7394}
\and H.~Kurki-Suonio\inst{\ref{aff71},\!\ref{aff72}}\orcidlink{0000-0002-4618-3063}
\and O.~Lahav\inst{\ref{aff6}}\orcidlink{0000-0002-1134-9035}
\and A.~M.~C.~Le~Brun\inst{\ref{aff73}}\orcidlink{0000-0002-0936-4594}
\and S.~Ligori\inst{\ref{aff5}}\orcidlink{0000-0003-4172-4606}
\and P.~B.~Lilje\inst{\ref{aff61}}\orcidlink{0000-0003-4324-7794}
\and V.~Lindholm\inst{\ref{aff71},\!\ref{aff72}}\orcidlink{0000-0003-2317-5471}\\
\and I.~Lloro\inst{\ref{aff74}}\orcidlink{0000-0001-5966-1434}
\and G.~Mainetti\inst{\ref{aff75}}\orcidlink{0000-0003-2384-2377}
\and D.~Maino\inst{\ref{aff76},\!\ref{aff30},\!\ref{aff77}}
\and E.~Maiorano\inst{\ref{aff13}}\orcidlink{0000-0003-2593-4355}
\and O.~Mansutti\inst{\ref{aff15}}\orcidlink{0000-0001-5758-4658}
\and S.~Marcin\inst{\ref{aff78}}
\and O.~Marggraf\inst{\ref{aff79}}\orcidlink{0000-0001-7242-3852}
\and K.~Markovic\inst{\ref{aff62}}\orcidlink{0000-0001-6764-073X}
\and M.~Martinelli\inst{\ref{aff31},\!\ref{aff32}}\orcidlink{0000-0002-6943-7732}
\and N.~Martinet\inst{\ref{aff27}}\orcidlink{0000-0003-2786-7790}
\and F.~Marulli\inst{\ref{aff80},\!\ref{aff13},\!\ref{aff19}}\orcidlink{0000-0002-8850-0303}
\and R.~Massey\inst{\ref{aff81}}\orcidlink{0000-0002-6085-3780}
\and S.~Maurogordato\inst{\ref{aff82}}
\and E.~Medinaceli\inst{\ref{aff13}}\orcidlink{0000-0002-4040-7783}
\and S.~Mei\inst{\ref{aff83},\!\ref{aff84}}\orcidlink{0000-0002-2849-559X}
\and M.~Melchior\inst{\ref{aff78}}
\and Y.~Mellier\inst{\ref{aff85},\!\ref{aff66}}\fnmsep\thanks{Deceased}\AAfnstore{deceased}
\and M.~Meneghetti\inst{\ref{aff13},\!\ref{aff19}}\orcidlink{0000-0003-1225-7084}
\and E.~Merlin\inst{\ref{aff31}}\orcidlink{0000-0001-6870-8900}
\and G.~Meylan\inst{\ref{aff86}}
\and A.~Mora\inst{\ref{aff87}}\orcidlink{0000-0002-1922-8529}
\and M.~Moresco\inst{\ref{aff80},\!\ref{aff13}}\orcidlink{0000-0002-7616-7136}
\and B.~Morin\inst{\ref{aff54}}
\and L.~Moscardini\inst{\ref{aff80},\!\ref{aff13},\!\ref{aff19}}\orcidlink{0000-0002-3473-6716}
\and E.~Munari\inst{\ref{aff15},\!\ref{aff14}}\orcidlink{0000-0002-1751-5946}
\and R.~Nakajima\inst{\ref{aff79}}\orcidlink{0009-0009-1213-7040}
\and C.~Neissner\inst{\ref{aff88},\!\ref{aff34}}\orcidlink{0000-0001-8524-4968}
\and R.~C.~Nichol\inst{\ref{aff10}}\orcidlink{0000-0003-0939-6518}
\and S.-M.~Niemi\inst{\ref{aff28}}
\and C.~Padilla\inst{\ref{aff88}}\orcidlink{0000-0001-7951-0166}
\and S.~Paltani\inst{\ref{aff51}}\orcidlink{0000-0002-8108-9179}
\and F.~Pasian\inst{\ref{aff15}}\orcidlink{0000-0002-4869-3227}
\and K.~Pedersen\inst{\ref{aff89}}
\and W.~J.~Percival\inst{\ref{aff90},\!\ref{aff91},\!\ref{aff92}}\orcidlink{0000-0002-0644-5727}
\and V.~Pettorino\inst{\ref{aff28}}
\and S.~Pires\inst{\ref{aff54}}\orcidlink{0000-0002-0249-2104}
\and G.~Polenta\inst{\ref{aff55}}\orcidlink{0000-0003-4067-9196}
\and M.~Poncet\inst{\ref{aff93}}
\and L.~A.~Popa\inst{\ref{aff94}}
\and L.~Pozzetti\inst{\ref{aff13}}\orcidlink{0000-0001-7085-0412}
\and F.~Raison\inst{\ref{aff58}}\orcidlink{0000-0002-7819-6918}
\and R.~Rebolo\inst{\ref{aff39},\!\ref{aff95},\!\ref{aff96}}
\and J.~Rhodes\inst{\ref{aff62}}\orcidlink{0000-0002-4485-8549}
\and G.~Riccio\inst{\ref{aff23}}
\and E.~Romelli\inst{\ref{aff15}}\orcidlink{0000-0003-3069-9222}
\and M.~Roncarelli\inst{\ref{aff13}}\orcidlink{0000-0001-9587-7822}
\and R.~Saglia\inst{\ref{aff60},\!\ref{aff58}}\orcidlink{0000-0003-0378-7032}
\and Z.~Sakr\inst{\ref{aff11},\!\ref{aff2},\!\ref{aff97}}\orcidlink{0000-0002-4823-3757}
\and D.~Sapone\inst{\ref{aff98}}\orcidlink{0000-0001-7089-4503}
\and B.~Sartoris\inst{\ref{aff60},\!\ref{aff15}}\orcidlink{0000-0003-1337-5269}
\and J.~A.~Schewtschenko\inst{\ref{aff40}}\orcidlink{0000-0002-4913-6393}
\and P.~Schneider\inst{\ref{aff79}}\orcidlink{0000-0001-8561-2679}
\and T.~Schrabback\inst{\ref{aff99}}\orcidlink{0000-0002-6987-7834}
\and A.~Secroun\inst{\ref{aff1}}\orcidlink{0000-0003-0505-3710}
\and E.~Sefusatti\inst{\ref{aff15},\!\ref{aff14},\!\ref{aff16}}\orcidlink{0000-0003-0473-1567}
\and G.~Seidel\inst{\ref{aff67}}\orcidlink{0000-0003-2907-353X}
\and S.~Serrano\inst{\ref{aff8},\!\ref{aff100},\!\ref{aff7}}\orcidlink{0000-0002-0211-2861}
\and P.~Simon\inst{\ref{aff79}}
\and C.~Sirignano\inst{\ref{aff101},\!\ref{aff53}}\orcidlink{0000-0002-0995-7146}
\and G.~Sirri\inst{\ref{aff19}}\orcidlink{0000-0003-2626-2853}
\and A.~Spurio~Mancini\inst{\ref{aff102}}\orcidlink{0000-0001-5698-0990}
\and L.~Stanco\inst{\ref{aff53}}\orcidlink{0000-0002-9706-5104}
\and J.-L.~Starck\inst{\ref{aff54}}\orcidlink{0000-0003-2177-7794}
\and J.~Steinwagner\inst{\ref{aff58}}\orcidlink{0000-0001-7443-1047}
\and P.~Tallada-Cresp\'{i}\inst{\ref{aff33},\!\ref{aff34}}\orcidlink{0000-0002-1336-8328}
\and D.~Tavagnacco\inst{\ref{aff15}}\orcidlink{0000-0001-7475-9894}
\and A.~N.~Taylor\inst{\ref{aff40}}
\and I.~Tereno\inst{\ref{aff49},\!\ref{aff103}}
\and S.~Toft\inst{\ref{aff104},\!\ref{aff105}}\orcidlink{0000-0003-3631-7176}
\and R.~Toledo-Moreo\inst{\ref{aff106}}\orcidlink{0000-0002-2997-4859}
\and F.~Torradeflot\inst{\ref{aff34},\!\ref{aff33}}\orcidlink{0000-0003-1160-1517}\\
\and J.~Valiviita\inst{\ref{aff71},\!\ref{aff72}}\orcidlink{0000-0001-6225-3693}
\and T.~Vassallo\inst{\ref{aff60},\!\ref{aff15}}\orcidlink{0000-0001-6512-6358}
\and G.~Verdoes~Kleijn\inst{\ref{aff107}}\orcidlink{0000-0001-5803-2580}
\and A.~Veropalumbo\inst{\ref{aff12},\!\ref{aff21},\!\ref{aff20}}\orcidlink{0000-0003-2387-1194}
\and Y.~Wang\inst{\ref{aff108}}\orcidlink{0000-0002-4749-2984}
\and J.~Weller\inst{\ref{aff60},\!\ref{aff58}}\orcidlink{0000-0002-8282-2010}
\and G.~Zamorani\inst{\ref{aff13}}\orcidlink{0000-0002-2318-301X}
\and F.~M.~Zerbi\inst{\ref{aff12}}
\and E.~Zucca\inst{\ref{aff13}}\orcidlink{0000-0002-5845-8132}
\and M.~Bolzonella\inst{\ref{aff13}}\orcidlink{0000-0003-3278-4607}
\and C.~Burigana\inst{\ref{aff109},\!\ref{aff56}}\orcidlink{0000-0002-3005-5796}
\and M.~Calabrese\inst{\ref{aff110},\!\ref{aff30}}\orcidlink{0000-0002-2637-2422}
\and D.~Di~Ferdinando\inst{\ref{aff19}}
\and J.~A.~Escartin~Vigo\inst{\ref{aff58}}
\and L.~Gabarra\inst{\ref{aff111}}\orcidlink{0000-0002-8486-8856}
\and S.~Matthew\inst{\ref{aff40}}\orcidlink{0000-0001-8448-1697}
\and N.~Mauri\inst{\ref{aff38},\!\ref{aff19}}\orcidlink{0000-0001-8196-1548}
\and A.~Pezzotta\inst{\ref{aff58}}\orcidlink{0000-0003-0726-2268}
\and M.~P\"ontinen\inst{\ref{aff71}}\orcidlink{0000-0001-5442-2530}
\and C.~Porciani\inst{\ref{aff79}}\orcidlink{0000-0002-7797-2508}
\and V.~Scottez\inst{\ref{aff85},\!\ref{aff112}}
\and M.~Tenti\inst{\ref{aff19}}\orcidlink{0000-0002-4254-5901}
\and M.~Viel\inst{\ref{aff14},\!\ref{aff15},\!\ref{aff17},\!\ref{aff16},\!\ref{aff113}}\orcidlink{0000-0002-2642-5707}
\and M.~Wiesmann\inst{\ref{aff61}}\orcidlink{0009-0000-8199-5860}
\and Y.~Akrami\inst{\ref{aff114},\!\ref{aff115}}\orcidlink{0000-0002-2407-7956}
\and V.~Allevato\inst{\ref{aff23}}\orcidlink{0000-0001-7232-5152}
\and I.~T.~Andika\inst{\ref{aff116},\!\ref{aff117}}\orcidlink{0000-0001-6102-9526}
\and M.~Archidiacono\inst{\ref{aff76},\!\ref{aff77}}\orcidlink{0000-0003-4952-9012}
\and F.~Atrio-Barandela\inst{\ref{aff118}}\orcidlink{0000-0002-2130-2513}
\and A.~Balaguera-Antolinez\inst{\ref{aff39},\!\ref{aff96}}\orcidlink{0000-0001-5028-3035}
\and M.~Ballardini\inst{\ref{aff119},\!\ref{aff13},\!\ref{aff120}}\orcidlink{0000-0003-4481-3559}
\and D.~Bertacca\inst{\ref{aff101},\!\ref{aff59},\!\ref{aff53}}\orcidlink{0000-0002-2490-7139}
\and M.~Bethermin\inst{\ref{aff121}}\orcidlink{0000-0002-3915-2015}
\and A.~Blanchard\inst{\ref{aff2}}\orcidlink{0000-0001-8555-9003}
\and L.~Blot\inst{\ref{aff122},\!\ref{aff73}}\orcidlink{0000-0002-9622-7167}
\and H.~B\"ohringer\inst{\ref{aff58},\!\ref{aff123},\!\ref{aff124}}\orcidlink{0000-0001-8241-4204}
\and S.~Borgani\inst{\ref{aff125},\!\ref{aff14},\!\ref{aff15},\!\ref{aff16},\!\ref{aff113}}\orcidlink{0000-0001-6151-6439}
\and M.~L.~Brown\inst{\ref{aff41}}\orcidlink{0000-0002-0370-8077}
\and S.~Bruton\inst{\ref{aff126}}\orcidlink{0000-0002-6503-5218}
\and R.~Cabanac\inst{\ref{aff2}}\orcidlink{0000-0001-6679-2600}
\and A.~Calabro\inst{\ref{aff31}}\orcidlink{0000-0003-2536-1614}
\and B.~Camacho~Quevedo\inst{\ref{aff8},\!\ref{aff7}}\orcidlink{0000-0002-8789-4232}
\and A.~Cappi\inst{\ref{aff13},\!\ref{aff82}}
\and F.~Caro\inst{\ref{aff31}}
\and C.~S.~Carvalho\inst{\ref{aff103}}
\and T.~Castro\inst{\ref{aff15},\!\ref{aff16},\!\ref{aff14},\!\ref{aff113}}\orcidlink{0000-0002-6292-3228}
\and F.~Cogato\inst{\ref{aff80},\!\ref{aff13}}\orcidlink{0000-0003-4632-6113}
\and S.~Contarini\inst{\ref{aff58}}\orcidlink{0000-0002-9843-723X}
\and T.~Contini\inst{\ref{aff2}}\orcidlink{0000-0003-0275-938X}
\and A.~R.~Cooray\inst{\ref{aff127}}\orcidlink{0000-0002-3892-0190}
\and S.~Davini\inst{\ref{aff21}}\orcidlink{0000-0003-3269-1718}
\and F.~De~Paolis\inst{\ref{aff128},\!\ref{aff129},\!\ref{aff130}}\orcidlink{0000-0001-6460-7563}
\and G.~Desprez\inst{\ref{aff107}}\orcidlink{0000-0001-8325-1742}
\and A.~D\'iaz-S\'anchez\inst{\ref{aff131}}\orcidlink{0000-0003-0748-4768}
\and S.~Di~Domizio\inst{\ref{aff20},\!\ref{aff21}}\orcidlink{0000-0003-2863-5895}
\and J.~M.~Diego\inst{\ref{aff132}}\orcidlink{0000-0001-9065-3926}
\and A.~G.~Ferrari\inst{\ref{aff19}}\orcidlink{0009-0005-5266-4110}
\and P.~G.~Ferreira\inst{\ref{aff111}}\orcidlink{0000-0002-3021-2851}
\and A.~Finoguenov\inst{\ref{aff71}}\orcidlink{0000-0002-4606-5403}
\and K.~Ganga\inst{\ref{aff83}}\orcidlink{0000-0001-8159-8208}
\and J.~Garc\'ia-Bellido\inst{\ref{aff114}}\orcidlink{0000-0002-9370-8360}
\and T.~Gasparetto\inst{\ref{aff15}}\orcidlink{0000-0002-7913-4866}
\and E.~Gaztanaga\inst{\ref{aff7},\!\ref{aff8},\!\ref{aff36}}\orcidlink{0000-0001-9632-0815}
\and F.~Giacomini\inst{\ref{aff19}}\orcidlink{0000-0002-3129-2814}
\and F.~Gianotti\inst{\ref{aff13}}\orcidlink{0000-0003-4666-119X}
\and G.~Gozaliasl\inst{\ref{aff133},\!\ref{aff71}}\orcidlink{0000-0002-0236-919X}
\and A.~Gregorio\inst{\ref{aff125},\!\ref{aff15},\!\ref{aff16}}\orcidlink{0000-0003-4028-8785}
\and M.~Guidi\inst{\ref{aff18},\!\ref{aff13}}\orcidlink{0000-0001-9408-1101}
\and C.~M.~Gutierrez\inst{\ref{aff134}}\orcidlink{0000-0001-7854-783X}
\and A.~Hall\inst{\ref{aff40}}\orcidlink{0000-0002-3139-8651}
\and S.~Hemmati\inst{\ref{aff135}}\orcidlink{0000-0003-2226-5395}
\and H.~Hildebrandt\inst{\ref{aff136}}\orcidlink{0000-0002-9814-3338}
\and J.~Hjorth\inst{\ref{aff89}}\orcidlink{0000-0002-4571-2306}
\and J.~J.~E.~Kajava\inst{\ref{aff137},\!\ref{aff138}}\orcidlink{0000-0002-3010-8333}
\and Y.~Kang\inst{\ref{aff51}}\orcidlink{0009-0000-8588-7250}
\and V.~Kansal\inst{\ref{aff139},\!\ref{aff140}}\orcidlink{0000-0002-4008-6078}
\and D.~Karagiannis\inst{\ref{aff119},\!\ref{aff141}}\orcidlink{0000-0002-4927-0816}
\and C.~C.~Kirkpatrick\inst{\ref{aff69}}
\and S.~Kruk\inst{\ref{aff43}}\orcidlink{0000-0001-8010-8879}
\and M.~Lattanzi\inst{\ref{aff120}}\orcidlink{0000-0003-1059-2532}
\and M.~Lembo\inst{\ref{aff119},\!\ref{aff120}}\orcidlink{0000-0002-5271-5070}
\and G.~Leroy\inst{\ref{aff142},\!\ref{aff81}}\orcidlink{0009-0004-2523-4425}
\and J.~Lesgourgues\inst{\ref{aff35}}\orcidlink{0000-0001-7627-353X}
\and T.~I.~Liaudat\inst{\ref{aff143}}\orcidlink{0000-0002-9104-314X}
\and S.~J.~Liu\inst{\ref{aff52}}\orcidlink{0000-0001-7680-2139}
\and A.~Loureiro\inst{\ref{aff144},\!\ref{aff145}}\orcidlink{0000-0002-4371-0876}
\and G.~Maggio\inst{\ref{aff15}}\orcidlink{0000-0003-4020-4836}
\and M.~Magliocchetti\inst{\ref{aff52}}\orcidlink{0000-0001-9158-4838}
\and F.~Mannucci\inst{\ref{aff146}}\orcidlink{0000-0002-4803-2381}
\and R.~Maoli\inst{\ref{aff147},\!\ref{aff31}}\orcidlink{0000-0002-6065-3025}
\and J.~Mart\'{i}n-Fleitas\inst{\ref{aff87}}\orcidlink{0000-0002-8594-569X}
\and C.~J.~A.~P.~Martins\inst{\ref{aff148},\!\ref{aff25}}\orcidlink{0000-0002-4886-9261}
\and L.~Maurin\inst{\ref{aff9}}\orcidlink{0000-0002-8406-0857}
\and R.~B.~Metcalf\inst{\ref{aff80},\!\ref{aff13}}\orcidlink{0000-0003-3167-2574}
\and M.~Miluzio\inst{\ref{aff43},\!\ref{aff149}}
\and P.~Monaco\inst{\ref{aff125},\!\ref{aff15},\!\ref{aff16},\!\ref{aff14}}\orcidlink{0000-0003-2083-7564}
\and C.~Moretti\inst{\ref{aff17},\!\ref{aff113},\!\ref{aff15},\!\ref{aff14},\!\ref{aff16}}\orcidlink{0000-0003-3314-8936}
\and C.~Murray\inst{\ref{aff83}}
\and S.~Nadathur\inst{\ref{aff36}}\orcidlink{0000-0001-9070-3102}
\and K.~Naidoo\inst{\ref{aff36}}\orcidlink{0000-0002-9182-1802}
\and A.~Navarro-Alsina\inst{\ref{aff79}}\orcidlink{0000-0002-3173-2592}
\and S.~Nesseris\inst{\ref{aff114}}\orcidlink{0000-0002-0567-0324}
\and K.~Paterson\inst{\ref{aff67}}\orcidlink{0000-0001-8340-3486}
\and A.~Pisani\inst{\ref{aff1},\!\ref{aff150}}\orcidlink{0000-0002-6146-4437}
\and D.~Potter\inst{\ref{aff151}}\orcidlink{0000-0002-0757-5195}
\and I.~Risso\inst{\ref{aff152}}\orcidlink{0000-0003-2525-7761}
\and P.-F.~Rocci\inst{\ref{aff9}}
\and M.~Sahl\'en\inst{\ref{aff153}}\orcidlink{0000-0003-0973-4804}
\and E.~Sarpa\inst{\ref{aff17},\!\ref{aff113},\!\ref{aff16}}\orcidlink{0000-0002-1256-655X}
\and A.~Schneider\inst{\ref{aff151}}\orcidlink{0000-0001-7055-8104}
\and D.~Sciotti\inst{\ref{aff31},\!\ref{aff32}}\orcidlink{0009-0008-4519-2620}
\and E.~Sellentin\inst{\ref{aff154},\!\ref{aff155}}
\and M.~Sereno\inst{\ref{aff13},\!\ref{aff19}}\orcidlink{0000-0003-0302-0325}
\and A.~Silvestri\inst{\ref{aff29}}\orcidlink{0000-0001-6904-5061}
\and L.~C.~Smith\inst{\ref{aff156}}\orcidlink{0000-0002-3259-2771}
\and K.~Tanidis\inst{\ref{aff111}}
\and C.~Tao\inst{\ref{aff1}}\orcidlink{0000-0001-7961-8177}
\and G.~Testera\inst{\ref{aff21}}
\and R.~Teyssier\inst{\ref{aff150}}\orcidlink{0000-0001-7689-0933}
\and S.~Tosi\inst{\ref{aff20},\!\ref{aff152}}\orcidlink{0000-0002-7275-9193}
\and A.~Troja\inst{\ref{aff101},\!\ref{aff53}}\orcidlink{0000-0003-0239-4595}
\and M.~Tucci\inst{\ref{aff51}}
\and C.~Valieri\inst{\ref{aff19}}
\and A.~Venhola\inst{\ref{aff157}}\orcidlink{0000-0001-6071-4564}
\and D.~Vergani\inst{\ref{aff13}}\orcidlink{0000-0003-0898-2216}
\and F.~Vernizzi\inst{\ref{aff158}}\orcidlink{0000-0003-3426-2802}
\and G.~Verza\inst{\ref{aff159}}\textcolor{blue}{\fnmsep\;\AAfnmark{deceased}}\orcidlink{0000-0002-1886-8348}
\and P.~Vielzeuf\inst{\ref{aff1}}\orcidlink{0000-0003-2035-9339}
\and N.~A.~Walton\inst{\ref{aff156}}\orcidlink{0000-0003-3983-8778}}

\institute{Aix-Marseille Universit\'e, CNRS/IN2P3, CPPM, Marseille, France\label{aff1}
\and
Institut de Recherche en Astrophysique et Plan\'etologie (IRAP), Universit\'e de Toulouse, CNRS, UPS, CNES, 14 Av. Edouard Belin, 31400 Toulouse, France\label{aff2}
\and
Dipartimento di Fisica, Universit\`a degli Studi di Torino, Via P. Giuria 1, 10125 Torino, Italy\label{aff3}\\
\and
INFN-Sezione di Torino, Via P. Giuria 1, 10125 Torino, Italy\label{aff4}
\and
INAF-Osservatorio Astrofisico di Torino, Via Osservatorio 20, 10025 Pino Torinese (TO), Italy\label{aff5}
\and
Department of Physics and Astronomy, University College London, Gower Street, London WC1E 6BT, UK\label{aff6}
\and
Institute of Space Sciences (ICE, CSIC), Campus UAB, Carrer de Can Magrans, s/n, 08193 Barcelona, Spain\label{aff7}
\and
Institut d'Estudis Espacials de Catalunya (IEEC),  Edifici RDIT, Campus UPC, 08860 Castelldefels, Barcelona, Spain\label{aff8}
\and
Universit\'e Paris-Saclay, CNRS, Institut d'astrophysique spatiale, 91405, Orsay, France\label{aff9}
\and
School of Mathematics and Physics, University of Surrey, Guildford, Surrey, GU2 7XH, UK\label{aff10}
\and
Institut f\"ur Theoretische Physik, University of Heidelberg, Philosophenweg 16, 69120 Heidelberg, Germany\label{aff11}
\and
INAF-Osservatorio Astronomico di Brera, Via Brera 28, 20122 Milano, Italy\label{aff12}
\and
INAF-Osservatorio di Astrofisica e Scienza dello Spazio di Bologna, Via Piero Gobetti 93/3, 40129 Bologna, Italy\label{aff13}
\and
IFPU, Institute for Fundamental Physics of the Universe, via Beirut 2, 34151 Trieste, Italy\label{aff14}
\and
INAF-Osservatorio Astronomico di Trieste, Via G. B. Tiepolo 11, 34143 Trieste, Italy\label{aff15}
\and
INFN, Sezione di Trieste, Via Valerio 2, 34127 Trieste TS, Italy\label{aff16}
\and
SISSA, International School for Advanced Studies, Via Bonomea 265, 34136 Trieste TS, Italy\label{aff17}
\and
Dipartimento di Fisica e Astronomia, Universit\`a di Bologna, Via Gobetti 93/2, 40129 Bologna, Italy\label{aff18}
\and
INFN-Sezione di Bologna, Viale Berti Pichat 6/2, 40127 Bologna, Italy\label{aff19}
\and
Dipartimento di Fisica, Universit\`a di Genova, Via Dodecaneso 33, 16146, Genova, Italy\label{aff20}
\and
INFN-Sezione di Genova, Via Dodecaneso 33, 16146, Genova, Italy\label{aff21}
\and
Department of Physics "E. Pancini", University Federico II, Via Cinthia 6, 80126, Napoli, Italy\label{aff22}
\and
INAF-Osservatorio Astronomico di Capodimonte, Via Moiariello 16, 80131 Napoli, Italy\label{aff23}
\and
INFN section of Naples, Via Cinthia 6, 80126, Napoli, Italy\label{aff24}
\and
Instituto de Astrof\'isica e Ci\^encias do Espa\c{c}o, Universidade do Porto, CAUP, Rua das Estrelas, PT4150-762 Porto, Portugal\label{aff25}
\and
Faculdade de Ci\^encias da Universidade do Porto, Rua do Campo de Alegre, 4150-007 Porto, Portugal\label{aff26}
\and
Aix-Marseille Universit\'e, CNRS, CNES, LAM, Marseille, France\label{aff27}
\and
European Space Agency/ESTEC, Keplerlaan 1, 2201 AZ Noordwijk, The Netherlands\label{aff28}
\and
Institute Lorentz, Leiden University, Niels Bohrweg 2, 2333 CA Leiden, The Netherlands\label{aff29}
\and
INAF-IASF Milano, Via Alfonso Corti 12, 20133 Milano, Italy\label{aff30}
\and
INAF-Osservatorio Astronomico di Roma, Via Frascati 33, 00078 Monteporzio Catone, Italy\label{aff31}
\and
INFN-Sezione di Roma, Piazzale Aldo Moro, 2 - c/o Dipartimento di Fisica, Edificio G. Marconi, 00185 Roma, Italy\label{aff32}
\and
Centro de Investigaciones Energ\'eticas, Medioambientales y Tecnol\'ogicas (CIEMAT), Avenida Complutense 40, 28040 Madrid, Spain\label{aff33}
\and
Port d'Informaci\'{o} Cient\'{i}fica, Campus UAB, C. Albareda s/n, 08193 Bellaterra (Barcelona), Spain\label{aff34}
\and
Institute for Theoretical Particle Physics and Cosmology (TTK), RWTH Aachen University, 52056 Aachen, Germany\label{aff35}
\and
Institute of Cosmology and Gravitation, University of Portsmouth, Portsmouth PO1 3FX, UK\label{aff36}
\and
Institute for Astronomy, University of Hawaii, 2680 Woodlawn Drive, Honolulu, HI 96822, USA\label{aff37}
\and
Dipartimento di Fisica e Astronomia "Augusto Righi" - Alma Mater Studiorum Universit\`a di Bologna, Viale Berti Pichat 6/2, 40127 Bologna, Italy\label{aff38}
\and
Instituto de Astrof\'{\i}sica de Canarias, V\'{\i}a L\'actea, 38205 La Laguna, Tenerife, Spain\label{aff39}
\and
Institute for Astronomy, University of Edinburgh, Royal Observatory, Blackford Hill, Edinburgh EH9 3HJ, UK\label{aff40}
\and
Jodrell Bank Centre for Astrophysics, Department of Physics and Astronomy, University of Manchester, Oxford Road, Manchester M13 9PL, UK\label{aff41}
\and
European Space Agency/ESRIN, Largo Galileo Galilei 1, 00044 Frascati, Roma, Italy\label{aff42}
\and
ESAC/ESA, Camino Bajo del Castillo, s/n., Urb. Villafranca del Castillo, 28692 Villanueva de la Ca\~nada, Madrid, Spain\label{aff43}
\and
Universit\'e Claude Bernard Lyon 1, CNRS/IN2P3, IP2I Lyon, UMR 5822, Villeurbanne, F-69100, France\label{aff44}
\and
Institut de Ci\`{e}ncies del Cosmos (ICCUB), Universitat de Barcelona (IEEC-UB), Mart\'{i} i Franqu\`{e}s 1, 08028 Barcelona, Spain\label{aff45}
\and
Instituci\'o Catalana de Recerca i Estudis Avan\c{c}ats (ICREA), Passeig de Llu\'{\i}s Companys 23, 08010 Barcelona, Spain\label{aff46}
\and
UCB Lyon 1, CNRS/IN2P3, IUF, IP2I Lyon, 4 rue Enrico Fermi, 69622 Villeurbanne, France\label{aff47}
\and
Mullard Space Science Laboratory, University College London, Holmbury St Mary, Dorking, Surrey RH5 6NT, UK\label{aff48}
\and
Departamento de F\'isica, Faculdade de Ci\^encias, Universidade de Lisboa, Edif\'icio C8, Campo Grande, PT1749-016 Lisboa, Portugal\label{aff49}
\and
Instituto de Astrof\'isica e Ci\^encias do Espa\c{c}o, Faculdade de Ci\^encias, Universidade de Lisboa, Campo Grande, 1749-016 Lisboa, Portugal\label{aff50}
\and
Department of Astronomy, University of Geneva, ch. d'Ecogia 16, 1290 Versoix, Switzerland\label{aff51}
\and
INAF-Istituto di Astrofisica e Planetologia Spaziali, via del Fosso del Cavaliere, 100, 00100 Roma, Italy\label{aff52}
\and
INFN-Padova, Via Marzolo 8, 35131 Padova, Italy\label{aff53}
\and
Universit\'e Paris-Saclay, Universit\'e Paris Cit\'e, CEA, CNRS, AIM, 91191, Gif-sur-Yvette, France\label{aff54}
\and
Space Science Data Center, Italian Space Agency, via del Politecnico snc, 00133 Roma, Italy\label{aff55}
\and
INFN-Bologna, Via Irnerio 46, 40126 Bologna, Italy\label{aff56}
\and
School of Physics, HH Wills Physics Laboratory, University of Bristol, Tyndall Avenue, Bristol, BS8 1TL, UK\label{aff57}
\and
Max Planck Institute for Extraterrestrial Physics, Giessenbachstr. 1, 85748 Garching, Germany\label{aff58}
\and
INAF-Osservatorio Astronomico di Padova, Via dell'Osservatorio 5, 35122 Padova, Italy\label{aff59}
\and
Universit\"ats-Sternwarte M\"unchen, Fakult\"at f\"ur Physik, Ludwig-Maximilians-Universit\"at M\"unchen, Scheinerstrasse 1, 81679 M\"unchen, Germany\label{aff60}
\and
Institute of Theoretical Astrophysics, University of Oslo, P.O. Box 1029 Blindern, 0315 Oslo, Norway\label{aff61}
\and
Jet Propulsion Laboratory, California Institute of Technology, 4800 Oak Grove Drive, Pasadena, CA, 91109, USA\label{aff62}
\and
Felix Hormuth Engineering, Goethestr. 17, 69181 Leimen, Germany\label{aff63}
\and
Technical University of Denmark, Elektrovej 327, 2800 Kgs. Lyngby, Denmark\label{aff64}
\and
Cosmic Dawn Center (DAWN), Denmark\label{aff65}
\and
Institut d'Astrophysique de Paris, UMR 7095, CNRS, and Sorbonne Universit\'e, 98 bis boulevard Arago, 75014 Paris, France\label{aff66}
\and
Max-Planck-Institut f\"ur Astronomie, K\"onigstuhl 17, 69117 Heidelberg, Germany\label{aff67}
\and
NASA Goddard Space Flight Center, Greenbelt, MD 20771, USA\label{aff68}
\and
Department of Physics and Helsinki Institute of Physics, Gustaf H\"allstr\"omin katu 2, 00014 University of Helsinki, Finland\label{aff69}\\
\and
Universit\'e de Gen\`eve, D\'epartement de Physique Th\'eorique and Centre for Astroparticle Physics, 24 quai Ernest-Ansermet, CH-1211 Gen\`eve 4, Switzerland\label{aff70}
\and
Department of Physics, P.O. Box 64, 00014 University of Helsinki, Finland\label{aff71}
\and
Helsinki Institute of Physics, Gustaf H{\"a}llstr{\"o}min katu 2, University of Helsinki, Helsinki, Finland\label{aff72}
\and
Laboratoire Univers et Th\'eorie, Observatoire de Paris, Universit\'e PSL, Universit\'e Paris Cit\'e, CNRS, 92190 Meudon, France\label{aff73}
\and
SKA Observatory, Jodrell Bank, Lower Withington, Macclesfield, Cheshire SK11 9FT, UK\label{aff74}
\and
Centre de Calcul de l'IN2P3/CNRS, 21 avenue Pierre de Coubertin 69627 Villeurbanne Cedex, France\label{aff75}
\and
Dipartimento di Fisica "Aldo Pontremoli", Universit\`a degli Studi di Milano, Via Celoria 16, 20133 Milano, Italy\label{aff76}
\and
INFN-Sezione di Milano, Via Celoria 16, 20133 Milano, Italy\label{aff77}
\and
University of Applied Sciences and Arts of Northwestern Switzerland, School of Engineering, 5210 Windisch, Switzerland\label{aff78}
\and
Universit\"at Bonn, Argelander-Institut f\"ur Astronomie, Auf dem H\"ugel 71, 53121 Bonn, Germany\label{aff79}
\and
Dipartimento di Fisica e Astronomia "Augusto Righi" - Alma Mater Studiorum Universit\`a di Bologna, via Piero Gobetti 93/2, 40129 Bologna, Italy\label{aff80}
\and
Department of Physics, Institute for Computational Cosmology, Durham University, South Road, Durham, DH1 3LE, UK\label{aff81}
\and
Universit\'e C\^{o}te d'Azur, Observatoire de la C\^{o}te d'Azur, CNRS, Laboratoire Lagrange, Bd de l'Observatoire, CS 34229, 06304 Nice cedex 4, France\label{aff82}
\and
Universit\'e Paris Cit\'e, CNRS, Astroparticule et Cosmologie, 75013 Paris, France\label{aff83}
\and
CNRS-UCB International Research Laboratory, Centre Pierre Binetruy, IRL2007, CPB-IN2P3, Berkeley, USA\label{aff84}
\and
Institut d'Astrophysique de Paris, 98bis Boulevard Arago, 75014, Paris, France\label{aff85}
\and
Institute of Physics, Laboratory of Astrophysics, Ecole Polytechnique F\'ed\'erale de Lausanne (EPFL), Observatoire de Sauverny, 1290 Versoix, Switzerland\label{aff86}
\and
Aurora Technology for European Space Agency (ESA), Camino bajo del Castillo, s/n, Urbanizacion Villafranca del Castillo, Villanueva de la Ca\~nada, 28692 Madrid, Spain\label{aff87}
\and
Institut de F\'{i}sica d'Altes Energies (IFAE), The Barcelona Institute of Science and Technology, Campus UAB, 08193 Bellaterra (Barcelona), Spain\label{aff88}
\and
DARK, Niels Bohr Institute, University of Copenhagen, Jagtvej 155, 2200 Copenhagen, Denmark\label{aff89}
\and
Waterloo Centre for Astrophysics, University of Waterloo, Waterloo, Ontario N2L 3G1, Canada\label{aff90}
\and
Department of Physics and Astronomy, University of Waterloo, Waterloo, Ontario N2L 3G1, Canada\label{aff91}
\and
Perimeter Institute for Theoretical Physics, Waterloo, Ontario N2L 2Y5, Canada\label{aff92}
\and
Centre National d'Etudes Spatiales -- Centre spatial de Toulouse, 18 avenue Edouard Belin, 31401 Toulouse Cedex 9, France\label{aff93}
\and
Institute of Space Science, Str. Atomistilor, nr. 409 M\u{a}gurele, Ilfov, 077125, Romania\label{aff94}
\and
Consejo Superior de Investigaciones Cientificas, Calle Serrano 117, 28006 Madrid, Spain\label{aff95}
\and
Universidad de La Laguna, Departamento de Astrof\'{\i}sica, 38206 La Laguna, Tenerife, Spain\label{aff96}
\and
Universit\'e St Joseph; Faculty of Sciences, Beirut, Lebanon\label{aff97}
\and
Departamento de F\'isica, FCFM, Universidad de Chile, Blanco Encalada 2008, Santiago, Chile\label{aff98}
\and
Universit\"at Innsbruck, Institut f\"ur Astro- und Teilchenphysik, Technikerstr. 25/8, 6020 Innsbruck, Austria\label{aff99}
\and
Satlantis, University Science Park, Sede Bld 48940, Leioa-Bilbao, Spain\label{aff100}
\and
Dipartimento di Fisica e Astronomia "G. Galilei", Universit\`a di Padova, Via Marzolo 8, 35131 Padova, Italy\label{aff101}
\and
Department of Physics, Royal Holloway, University of London, TW20 0EX, UK\label{aff102}
\and
Instituto de Astrof\'isica e Ci\^encias do Espa\c{c}o, Faculdade de Ci\^encias, Universidade de Lisboa, Tapada da Ajuda, 1349-018 Lisboa, Portugal\label{aff103}
\and
Cosmic Dawn Center (DAWN)\label{aff104}
\and
Niels Bohr Institute, University of Copenhagen, Jagtvej 128, 2200 Copenhagen, Denmark\label{aff105}
\and
Universidad Polit\'ecnica de Cartagena, Departamento de Electr\'onica y Tecnolog\'ia de Computadoras,  Plaza del Hospital 1, 30202 Cartagena, Spain\label{aff106}
\and
Kapteyn Astronomical Institute, University of Groningen, PO Box 800, 9700 AV Groningen, The Netherlands\label{aff107}
\and
Infrared Processing and Analysis Center, California Institute of Technology, Pasadena, CA 91125, USA\label{aff108}
\and
INAF, Istituto di Radioastronomia, Via Piero Gobetti 101, 40129 Bologna, Italy\label{aff109}
\and
Astronomical Observatory of the Autonomous Region of the Aosta Valley (OAVdA), Loc. Lignan 39, I-11020, Nus (Aosta Valley), Italy\label{aff110}
\and
Department of Physics, Oxford University, Keble Road, Oxford OX1 3RH, UK\label{aff111}
\and
ICL, Junia, Universit\'e Catholique de Lille, LITL, 59000 Lille, France\label{aff112}
\and
ICSC - Centro Nazionale di Ricerca in High Performance Computing, Big Data e Quantum Computing, Via Magnanelli 2, Bologna, Italy\label{aff113}
\and
Instituto de F\'isica Te\'orica UAM-CSIC, Campus de Cantoblanco, 28049 Madrid, Spain\label{aff114}
\and
CERCA/ISO, Department of Physics, Case Western Reserve University, 10900 Euclid Avenue, Cleveland, OH 44106, USA\label{aff115}
\and
Technical University of Munich, TUM School of Natural Sciences, Physics Department, James-Franck-Str.~1, 85748 Garching, Germany\label{aff116}
\and
Max-Planck-Institut f\"ur Astrophysik, Karl-Schwarzschild-Str.~1, 85748 Garching, Germany\label{aff117}
\and
Departamento de F{\'\i}sica Fundamental. Universidad de Salamanca. Plaza de la Merced s/n. 37008 Salamanca, Spain\label{aff118}
\and
Dipartimento di Fisica e Scienze della Terra, Universit\`a degli Studi di Ferrara, Via Giuseppe Saragat 1, 44122 Ferrara, Italy\label{aff119}
\and
Istituto Nazionale di Fisica Nucleare, Sezione di Ferrara, Via Giuseppe Saragat 1, 44122 Ferrara, Italy\label{aff120}
\and
Universit\'e de Strasbourg, CNRS, Observatoire astronomique de Strasbourg, UMR 7550, 67000 Strasbourg, France\label{aff121}
\and
Center for Data-Driven Discovery, Kavli IPMU (WPI), UTIAS, The University of Tokyo, Kashiwa, Chiba 277-8583, Japan\label{aff122}
\and
Ludwig-Maximilians-University, Schellingstrasse 4, 80799 Munich, Germany\label{aff123}
\and
Max-Planck-Institut f\"ur Physik, Boltzmannstr. 8, 85748 Garching, Germany\label{aff124}
\and
Dipartimento di Fisica - Sezione di Astronomia, Universit\`a di Trieste, Via Tiepolo 11, 34131 Trieste, Italy\label{aff125}
\and
California Institute of Technology, 1200 E California Blvd, Pasadena, CA 91125, USA\label{aff126}
\and
Department of Physics \& Astronomy, University of California Irvine, Irvine CA 92697, USA\label{aff127}
\and
Department of Mathematics and Physics E. De Giorgi, University of Salento, Via per Arnesano, CP-I93, 73100, Lecce, Italy\label{aff128}
\and
INFN, Sezione di Lecce, Via per Arnesano, CP-193, 73100, Lecce, Italy\label{aff129}
\and
INAF-Sezione di Lecce, c/o Dipartimento Matematica e Fisica, Via per Arnesano, 73100, Lecce, Italy\label{aff130}
\and
Departamento F\'isica Aplicada, Universidad Polit\'ecnica de Cartagena, Campus Muralla del Mar, 30202 Cartagena, Murcia, Spain\label{aff131}
\and
Instituto de F\'isica de Cantabria, Edificio Juan Jord\'a, Avenida de los Castros, 39005 Santander, Spain\label{aff132}
\and
Department of Computer Science, Aalto University, PO Box 15400, Espoo, FI-00 076, Finland\label{aff133}
\and
Instituto de Astrof\'\i sica de Canarias, c/ Via Lactea s/n, La Laguna 38200, Spain. Departamento de Astrof\'\i sica de la Universidad de La Laguna, Avda. Francisco Sanchez, La Laguna, 38200, Spain\label{aff134}
\and
Caltech/IPAC, 1200 E. California Blvd., Pasadena, CA 91125, USA\label{aff135}
\and
Ruhr University Bochum, Faculty of Physics and Astronomy, Astronomical Institute (AIRUB), German Centre for Cosmological Lensing (GCCL), 44780 Bochum, Germany\label{aff136}
\and
Department of Physics and Astronomy, Vesilinnantie 5, 20014 University of Turku, Finland\label{aff137}
\and
Serco for European Space Agency (ESA), Camino bajo del Castillo, s/n, Urbanizacion Villafranca del Castillo, Villanueva de la Ca\~nada, 28692 Madrid, Spain\label{aff138}
\and
ARC Centre of Excellence for Dark Matter Particle Physics, Melbourne, Australia\label{aff139}
\and
Centre for Astrophysics \& Supercomputing, Swinburne University of Technology,  Hawthorn, Victoria 3122, Australia\label{aff140}
\and
Department of Physics and Astronomy, University of the Western Cape, Bellville, Cape Town, 7535, South Africa\label{aff141}
\and
Department of Physics, Centre for Extragalactic Astronomy, Durham University, South Road, Durham, DH1 3LE, UK\label{aff142}
\and
IRFU, CEA, Universit\'e Paris-Saclay 91191 Gif-sur-Yvette Cedex, France\label{aff143}
\and
Oskar Klein Centre for Cosmoparticle Physics, Department of Physics, Stockholm University, Stockholm, SE-106 91, Sweden\label{aff144}
\and
Astrophysics Group, Blackett Laboratory, Imperial College London, London SW7 2AZ, UK\label{aff145}
\and
INAF-Osservatorio Astrofisico di Arcetri, Largo E. Fermi 5, 50125, Firenze, Italy\label{aff146}
\and
Dipartimento di Fisica, Sapienza Universit\`a di Roma, Piazzale Aldo Moro 2, 00185 Roma, Italy\label{aff147}
\and
Centro de Astrof\'{\i}sica da Universidade do Porto, Rua das Estrelas, 4150-762 Porto, Portugal\label{aff148}
\and
HE Space for European Space Agency (ESA), Camino bajo del Castillo, s/n, Urbanizacion Villafranca del Castillo, Villanueva de la Ca\~nada, 28692 Madrid, Spain\label{aff149}
\and
Department of Astrophysical Sciences, Peyton Hall, Princeton University, Princeton, NJ 08544, USA\label{aff150}
\and
Department of Astrophysics, University of Zurich, Winterthurerstrasse 190, 8057 Zurich, Switzerland\label{aff151}
\and
INAF-Osservatorio Astronomico di Brera, Via Brera 28, 20122 Milano, Italy, and INFN-Sezione di Genova, Via Dodecaneso 33, 16146, Genova, Italy\label{aff152}
\and
Theoretical astrophysics, Department of Physics and Astronomy, Uppsala University, Box 515, 751 20 Uppsala, Sweden\label{aff153}
\and
Mathematical Institute, University of Leiden, Einsteinweg 55, 2333 CA Leiden, The Netherlands\label{aff154}
\and
Leiden Observatory, Leiden University, Einsteinweg 55, 2333 CC Leiden, The Netherlands\label{aff155}
\and
Institute of Astronomy, University of Cambridge, Madingley Road, Cambridge CB3 0HA, UK\label{aff156}
\and
Space physics and astronomy research unit, University of Oulu, Pentti Kaiteran katu 1, FI-90014 Oulu, Finland\label{aff157}
\and
Institut de Physique Th\'eorique, CEA, CNRS, Universit\'e Paris-Saclay 91191 Gif-sur-Yvette Cedex, France\label{aff158}
\and
Center for Computational Astrophysics, Flatiron Institute, 162 5th Avenue, 10010, New York, NY, USA\label{aff159}}       

\date{Received 14 March 2025 / Accepted 26 May 2026}

\abstract{
With about 1.5 billion galaxies expected to be observed, the very large number of objects in the \Euclid photometric survey will allow for precise studies of galaxy clustering from a single survey, over a large range of redshifts, $0.2 < z < 2.5$. In this work, we use photometric redshifts \zph\,to extract the baryon acoustic oscillation signal (BAO) from the Flagship galaxy mock catalogue with a tomographic approach to constrain the evolution of the Universe and infer its cosmological parameters. We measured the two-point angular correlation function in 13 redshift bins. A template-fitting approach was applied to the measurement to extract the shift of the BAO peak through the transverse Alcock--Paczynski parameter $\alpha$. A joint analysis of all redshift bins was performed to constrain $\alpha$ at the effective redshift $z_\mathrm{eff}=0.77$ with Markov chain Monte Carlo and profile likelihood techniques. We also extracted one $\alpha_i$ parameter per redshift bin to quantify its evolution as a function of time. From these 13 $\alpha_i$, which are directly proportional to the ratio $D_\mathrm{A}/r_\mathrm{d}$, we constrain the product of the reduced Hubble constant and the sound horizon at the drag epoch, $h\,r_\mathrm{d}$, and the matter density parameter $\Omm$. From the joint analysis, we constrain $D_\mathrm{A}/r_\mathrm{d}=10.764^{+0.0839}_{-0.0849}$ at the 68\% confidence level, which represents a three-fold improvement over current constraints from the Dark Energy Survey. As expected, the constraining power in the analysis of each redshift bin is lower, with an uncertainty ranging from $\pm1.23$ to $\pm0.289$. From these results, we constrain $\Omm=0.296^{+0.074}_{-0.059}$ and $h\,r_\mathrm{d}=99.35^{+4.29}_{-4.55}\,\mathrm{Mpc}$. We quantify the influence of analysis choices such as the template, scale cuts, and redshift bins, and systematic effects such as redshift-space distortions, over our constraints, both at the level of the extracted $\alpha_i$ parameters and at the level of cosmological inference.
}

\keywords{cosmological parameters -- cosmology: theory -- large-scale structure of Universe}

\titlerunning{\aap, 713, A281 (2026)}
\authorrunning{Euclid Collaboration: Duret et al.}
   
\maketitle

\nolinenumbers
\section{Introduction}
\label{sc:Intro}
\vspace{0.1cm}
As \;a \;stage \;IV \;survey, \,\Euclid \,\citep{Laureijs_11,Mellier_2025} was primarily designed to constrain dark energy with two main probes: weak lensing and spectroscopic galaxy clustering. The former will make use of galaxy shapes observed with the Visible Camera \citep[VIS,][]{Cropper_2025} and their photometric redshifts obtained with the photometer of the Near-Infrared Spectrometer and Photometer \citep[NISP,][]{Jahnke_2025} together with ground-based observations. The latter will use the precise measurements of galaxy redshifts obtained with the spectrometer of the NISP. \Euclid will provide a photometric sample of about 1.5 billion galaxies that can be used not only for weak lensing, but also for many other probes such as photometric galaxy clustering. The combination of weak gravitational lensing with photometric galaxy clustering will provide strong cosmological constraints \citep{Blanchard_2020,Tutusaus2020}, which motivates considering this probe in addition to the standard spectroscopic galaxy clustering.

The clustering of galaxies puts constraints on the expansion history of the Universe. One of its most constraining features is the size of the baryon acoustic oscillations (BAOs). The BAO scale is a characteristic scale of the Universe that corresponds to the imprint left in the distribution of galaxies by primordial oscillations of the baryons when they were still coupled to photons. These oscillations were created by the interplay between the radiation pressure force supported by photons and the gravitational pull of dark matter overdensities. When baryons and photons decoupled at the drag epoch, oscillations stopped and froze at a scale known as the BAO scale. It can be observed as a peak in the correlation function of the galaxy density field or a succession of peaks in its power spectrum. While it is fixed in\, comoving\, coordinates,\, the\, apparent\, size\, of\, the\, BAO\, scale\, increases as the Universe expands so that constraining this scale at different redshifts provides information on the expansion rate of the Universe.

\vspace{-0.05cm}The \,BAO \,signal \,is \,traditionally \,constrained \,in \,3D \,with spectroscopic \,redshifts \,but \,given \,the \,current \,accuracy \,and \,precision of photometric redshifts, useful information from the BAO \;signal \;can \;also \;be \;extracted \;using \;photometric \;samples. The first observations of the \,BAO \,signal \,in \,galaxy \,surveys were performed with the Sloan Digital Sky Survey \citep{Eisenstein_2005} and the 2-degree Field Galaxy Redshift Survey \mbox{\citep{Percival_2001,Cole_2005}} and recently reached new levels of precision with the 0.52\% joint constraints on BAO obtained by the Dark Energy Spectroscopic Instrument first year of observations \citep{Adame_2025} and with constraints varying between 2.1\% and 7.3\% obtained on individual\, bins\, in\, its\, second\, data\, release\, \citep[no joint constraint on $\alpha$]{Abdul_Karim_2025}. Spectroscopic redshift measurements are\, very\, accurate\, but\, not\, fast\, enough\, to\, obtain\, the\, redshift\, of all galaxies detected in the photometric sample. \Euclid uses slitless spectroscopy, allowing for multiple spectra to be measured in a single exposure. However, the mission optimisation resulted in \Euclid being capable of reliably detecting emission lines down to a flux limit of $2\expo{-16}\;\mathrm{erg}\,\mathrm{s}^{-1}\,\mathrm{cm}^{-2}$ in its wide survey \citep{Mellier_2025}. On the contrary, photometric redshifts are obtained from multi-band wide filters instead of spectra so they can be measured for all the galaxies of the photometric sample. Despite their lower accuracy, they are now available for such a large number of galaxies that they can be used to put significant constraints on cosmology. One recent example is the latest results from the Dark Energy Survey (DES), which constrain the BAO shift parameter $\alpha$ with an uncertainty of 2.1\% at $z_\mathrm{eff}=0.85$ \citep[][DES Y6 from now on]{Abbott_2024_DESY6} with a sample of almost 16 million galaxies selected among \,over \,300 \,million \,observed. To do so, a tomographic approach is typically used, dividing the full galaxy sample into redshift bins and measuring the angular correlation function or angular power spectrum in each bin. In \citetalias{Abbott_2024_DESY6}, six bins were defined between $0.6 < \zph < 1.2$. In this context, the \Euclid photometric survey can increase its constraining power by including its photometric sample for galaxy clustering studies in addition to weak lensing analyses. One advantage of \Euclid is that it will provide numerous photometric redshifts (about $1.5 \expo{9}$) in a much larger redshift range than was previously available, covering $0.2 < \zph < 2.5$.

In this work, we study how well the \Euclid photometric sample will be able to constrain the BAO scale, using the Flagship simulation \citep{Castander_2025}. We consider two analyses: one in which we extract the BAO scale in each of the 13 redshift bins, yielding 13 values of $\alpha$ and thus the evolution of the parameter as a function of redshift, and a second in which we conduct a joint analysis of all the redshift bins to constrain a single value of $\alpha$. We adopt the 13 redshift bins described in \citet{Mellier_2025} as this was shown to optimise \Euclid Data Release 3 (DR3) analysis results \citep{Pocino2021}. Here, we focus on the two-point angular correlation function in configuration space $w(\theta)$ as an observable to detect the BAO signal in the Flagship galaxy mock catalogue. While \Euclid is expected to cover an area of about $14\,000$ deg$^2$ \citep{Mellier_2025}, the Flagship simulation covers 37\% of this area. This means that the area used in this work is intermediate between DR1 and DR2 for \Euclid, which are expected to cover approximately 2500 and 7500 deg$^2$, respectively. We want to quantify our ability to constrain this signal as well as estimate the weight of different choices in the setup of the analysis.

The paper is organised as follows. In Sect.~\ref{sc:Data}, we present the data from the Flagship simulation used throughout this work. We describe the theoretical framework of the analysis in Sect.~\ref{sc:Theory} with the computation of the two-point angular correlation function model, its estimator, and its covariance. The method used to extract the BAO scale and to infer cosmological parameters is detailed in Sect.~\ref{sc:methodology}. Results are reported and discussed in Sect.~\ref{sc:Results}, including our joint analysis, the individual redshift bin analysis, and cosmological constraints along with their robustness to choices of fitting templates, scale cuts, redshift-space distortions (RSDs), and redshift binning scheme. We present our main conclusions in Sect.~\ref{sc:Conclusions}.

\section{Data}
\label{sc:Data}

We use the Flagship v.2.1.10 galaxy mock sample \citep{Castander_2025} available on CosmoHub\,\footnote{\url{https://cosmohub.pic.es/catalogs/353}} \citep{TALLADA2020100391,Carretero_2017} created from the Flagship N-body dark matter simulation \citep{Potter_2017}. This simulation assumes the following flat $\Lambda$ cold dark matter ($\Lambda$CDM) cosmology: matter density parameter $\Omm = 0.319$, baryon density parameter $\Omb = 0.049$, dark energy density parameter $\Omega_{\Lambda} = 0.681 - \Omega_\mathrm{r} - \Omega_{\nu}$, with a radiation density parameter $\Omega_\mathrm{r} = 0.00005509$ and $\Omega_{\nu} = 0.00140343$ for massive neutrinos, dark energy equation of state parameter $w_\mathrm{de} = -1.0$, reduced Hubble constant $h = 0.67$, spectral index of the primordial power spectrum $n_\mathrm{s} = 0.96$, and its amplitude $A_\mathrm{s} = 2.1 \expo{-9}$ at $k = 0.05 \,\rm Mpc^{-1}$. This simulation considers a $3.6\,h^{-1}\,\rm Gpc$-side box with $4\expo{12}$ particles of mass $10^9\,h^{-1}M_\odot$. The main output of the simulation is a lightcone that spans redshifts between $0$ and $3$. Dark matter haloes were identified down to $10^{10}\,h^{-1}\,M_\odot$ with \texttt{ROCKSTAR}\,\citep{Behroozi_2013}. These haloes were then populated with galaxies with halo abundance matching and halo occupation distribution techniques following \citet{Carretero_2015}. Galaxy luminosities were calibrated with a combination of the luminosity functions from \citet{Blanton_2003}, \citet{Blanton_2005b}, and \citet{Dahlen_2005}. Galaxy clustering measurements were calibrated as a function of colour and luminosity following \citet{Zehavi_2011} and the colour-magnitude diagram from \citet{Blanton_2005a} was used as a reference.

\begin{figure}
    \centering
    \resizebox{\hsize}{!}{\includegraphics{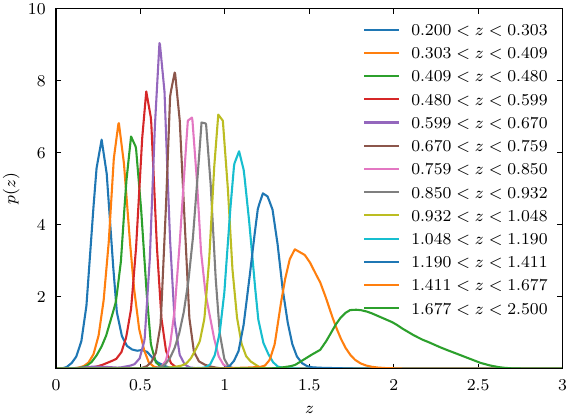}}
    \caption{True redshift distribution of the galaxies from Flagship 2.1 \\selected in 13 equipopulated photometric redshift bins.}
    \label{fig:flagship_nz}
\end{figure}

We applied a magnitude cut at $\IE \leq 24.5$ in the VIS \IE band with the additional constraint that only objects with properly determined photometric redshifts are considered (\texttt{phz\_flags = 0}). This sample covers one octant of the sky between right ascension $145^\circ \leq \mathrm{RA} < 235^\circ$ and declination $0^\circ< \mathrm{Dec} < 90^\circ$ for a total area of $5157\,\mathrm{deg}^2$. Following \citet{Mellier_2025}, we divided the sample into 13 equipopulated redshift bins of \zph. The normalised redshift distributions $p_i(z)$ of the 13 bins are shown in Fig.~\ref{fig:flagship_nz}. This division yields a large statistical sample, with about $34.8$ million galaxies per redshift bin. At the end of Sect.~\ref{sc:results_individual_bins}, we also consider a different sample used for a DR1-like setting with six equipopulated redshift bins, a selection cut at $\IE\leq23.5$, and covering 2500 deg$^2$. The resulting redshift bin limits are 0.2, 0.396, 0.507, 0.657, 0.84, 1.04, and 2.5 with about 10.8 million galaxies per bin.

Photometric redshifts are defined as the first mode of the probability density functions (PDFs) obtained with the k-nearest neighbours algorithm \texttt{NNPZ} \citep{Tanaka_2018} by matching galaxy magnitude and colours to a reference sample of 2 million galaxies simulated up to the depth of the \Euclid calibration field $\IE=29.4$ and whose PDFs are obtained using the template-fitting code \texttt{Phosphoros} \citep{Tucci_2025}. The photometric redshift PDF of each galaxy is computed as a weighted average of the PDF of the neighbours found by \texttt{NNPZ}, the weight being the inverse of the $\chi^2$ distance between the galaxy and the neighbour in the magnitude and colour space. The constraint \texttt{phz\_flags = 0} ensures that the galaxy had enough neighbours found to properly derive the photometric redshift when \texttt{NNPZ} was applied. The photometry used to infer these redshifts has the level of noise expected from the depth of the ground-based observations of the Legacy Survey of Space and Time \citep[LSST,][]{Ivezic_2019} for all galaxies, which is optimistic ($u=24.4$, $g=25.6$, $r=25.7$, $i=25.0$, $z=24.3$, $y=24.3$). We used the normalised mean absolute deviation to compare the quality of these photometric redshifts, 
\begin{equation}
\sigma_\mathrm{NMAD} = 1.4826\,\mathrm{median}\left(\left|\Delta z-\mathrm{median  }\left(\Delta z\right)\right|\right),
\label{eq:sigma_nmad}
\end{equation}
where $\Delta z=(\zph-z_\mathrm{true})(1+z_\mathrm{true})^{-1}$. The complete table of $\sigma_\mathrm{NMAD}$ and other metrics is given for all redshift bins in Appendix~\ref{ap:zph_stats} and shown in Fig.~\ref{fig:photo_zs_stats_grouped}. The fiducial cosmology used in this paper is a flat $\Lambda$CDM cosmology, defined by a set of parameters that match the fiducial cosmology of the Flagship simulation.

\section{Two-point angular correlation function}
\label{sc:Theory}

In this section, we describe the observable relevant to galaxy clustering with the photometric survey that is considered in this analysis: the galaxy two-point angular correlation function. We also present its estimator and its covariance.

\subsection{Theory}
\label{sc:theory_2pcf}
As we process information projected into bins of redshift, we consider the galaxy two-point angular correlation function $w(\theta)$, defined as \citep[e.g.][]{Crocce_2011}
\begin{equation}
w(\theta)=\sum\limits_{\ell\geq 0}\frac{2\ell+1}{4\pi}\,C(\ell)\,P_\ell(\cos\,\theta),
\label{eq:w_theory}
\end{equation}
with $P_\ell$ being the Legendre polynomial and $C(\ell)$ the angular power spectrum defined as
\begin{equation}
C(\ell) = 4\pi\int_0^{\infty}\frac{\diff k}{k}\,\mathcal{P}_{\Phi}(k)\;\Delta^2_{\ell}(k) .
\label{Cl_theory}
\end{equation}
In Eq.~\eqref{Cl_theory}, $k$ is the wavenumber and $\mathcal{P}_{\Phi}$ stands for the dimensionless power spectrum of the primordial curvature perturbations $\Phi({k})$ that we model with \texttt{HMCode} \citep{Mead_2021} as implemented in the Boltzmann code \texttt{CAMB} \citep{Challinor_2011}. The term $\Delta_{\ell}$ is the sum of the transfer function of number counts for the galaxy density field, $\Delta^{\rm D}_{\ell}$, and the linear RSD contribution, $\Delta^{\rm RSD}_{\ell}$ \citep{Chisari_2019}:
\begin{equation}
\Delta_{\ell}(k) = \Delta^{\rm D}_{\ell}(k) + \Delta^{\rm RSD}_{\ell}(k),
\end{equation}
where the two terms are defined, respectively, as
\begin{equation}
\Delta^{\rm D}_{\ell}(k) := \int_0^{z_{\rm max}} \diff z\;p(z)\;b(z)\;T_{\delta}(k,z)\;j_{\ell}\left(kr\right)
\label{density_field_theory}
\end{equation}
and 
\begin{equation}
\Delta^{\rm RSD}_{\ell}(k) := \int_0^{z_{\rm max}} \diff z\;\frac{(1+z)\,p(z)}{H(z)}T_{\theta}(k,z)\;j^{\prime\prime}_{\ell}\left(kr\right),
\label{rsd_theory}
\end{equation}
where $p(z)$ is the normalised galaxy redshift distribution, $b(z)$ is the linear galaxy bias, $z_{\rm max}$ is the maximum redshift of the survey that we fix to 3 because $p(z\geq3) = 0$, $j_\ell$ is the spherical Bessel function of order $\ell$, $j_\ell''$ its second derivative, and $r(z)$ is the comoving radial distance. In Eqs.~\eqref{density_field_theory} and \eqref{rsd_theory}, the transfer function $T_\mathrm{X}(k,z)$ of a quantity, $X$, is defined as the ratio $X(k,z)/\Phi(k)$ so that $T_{\delta}$ is the transfer function of the matter overdensity $\delta(k,z)$ and $T_{\theta}$ is the transfer function of the divergence of the comoving velocity field. Discussed in \citet{Lepori_2022} regarding full-shape analyses of photometric galaxy clustering, we checked that the effect of magnification bias on $\alpha$ is smaller than $0.2\,\sigma$ in most redshift bins, which is why no $\Delta^{\rm M}_{\ell}(k)$ term as defined in Eq.~(26) of \citet{Chisari_2019} is included in the model. Other relativistic effects are completely sub-dominant and not included either \citep[see Eq.~10 and Appendix~A of][]{Alonso_2015}.

To accurately model large scales at $\ell<220$ needed for BAO analysis, non-Limber integrals were computed using the Fang--Krause--Eifler--MacCrann (FKEM) method described in \citet{Fang_2020}, while the faster Limber approximation \citep{Kaiser_1992} was used at $\ell\in[220,10\,000]$. We checked that the effect of the Limber approximation on $C(\ell)$ is smaller than 1\% in this range of $\ell$. Throughout this work, the theoretical $w(\theta)$ was computed using the Core Cosmology Library \citep{Chisari_2019}. We refer to $w_\mathrm{fid}(\theta)$ when the theoretical $w(\theta)$ was computed with the fiducial cosmological parameters defined in Sect.~\ref{sc:Data}.

\subsection{Estimator}
\label{sc:data_2pcf}

The two-point angular correlation function was computed from the Flagship galaxy mock catalogue (see Sect.~\ref{sc:Data}) with the Landy--Szalay estimator \citep{Landy_1993}
\begin{equation}
w(\theta)=\frac{\NDD-2\NDR+\NRR}{\NRR},
\end{equation}
where \NDD, \NDR, and \NRR\, are the normalised pair counts at angular separation $\theta$ where D stands for data and R for a random point. The random catalogues were created by sampling the footprint of the Flagship simulation defined by a \texttt{HEALPix} mask of $N_\mathrm{side} = 4096$, which is equivalent to an angular resolution of $0\degf014$. We used 30 times as many random points as galaxies, yielding $1.04\expo{9}$ points per redshift bin. The angular binning has a resolution of $0\degf1$ and spans between $\thmin=0\degf12$ \citep[intermediate between the optimistic and pessimistic scenarios considered in][]{Blanchard_2020} and $\thmax=\thbao+2\degf5$, with 
\begin{equation}
\thbao = \frac{r_\mathrm{d,\,fid}}{(1+z_\mathrm{eff})\;D_\mathrm{A,\,fid}(z_\mathrm{eff})},
\end{equation}
evaluated at the mean redshift of the bin $z_\mathrm{eff}$ for the cosmology of the simulation given in Sect.~\ref{sc:Data}. The measurement of $w(\theta)$ was performed with the code \texttt{TreeCorr} \citep{Jarvis_2004}.

\subsection{Covariance}
\label{sc:covariance}
Two approaches were considered to compute the covariance of the two-point angular correlation function. For individual bin analyses, in which one value of the $\alpha$ parameter is fitted for each redshift bin, we used the jackknife covariance matrix built from the data vector measured with \texttt{TreeCorr}, while joint analyses used an analytical covariance computed with \texttt{CosmoCov} \citep{Krause_2017}. The jackknife covariance was computed by resampling $N_\mathrm{patch}=500$ patches of approximately $10\, \mathrm{deg}^2$ each. This choice was made to ensure that the jackknife uncertainties converged with respect to the number of patches. For instance, using $N_\mathrm{patch}=210$  yields jackknife errors that are nearly 20\% larger than those obtained with $N_\mathrm{patch}=500$ on large scales, whereas increasing the number from 500 to 1000 patches changes the errors by only $\sim$\,5\%. The use of an analytical covariance was made necessary by the noise in the jackknife covariance of the 13 redshift bins, even with 500 patches. Increasing to a larger number of patches yields very little improvement in that regard.

The jackknife covariance matrix was computed with
\begin{multline}
\mathrm{Cov}^\mathrm{JK}(w_{ij}\,(\theta),w_{kl}\,(\theta')) \\ = \frac{N_\mathrm{patch}-1}{N_\mathrm{patch}}\sum_{n=1}^{N_\mathrm{patch}}\left(w_{ij}^{(n)}\,(\theta)-\overline{w}_{ij}\,(\theta)\right)\left(w_{kl}^{(n)}\,(\theta')-\overline{w}_{kl}\,(\theta')\right),
\label{eq:cov_jackknife}
\end{multline}
where $n$ is the index of the jackknife realisation, $w_{ij}\,(\theta)$ is the correlation of redshift bins $i$ and $j$ at angular separation $\theta$, and $\overline{w}_{ij}$ stands for the mean over all jackknife realisations for each angular separation $\overline{w}_{ij}\,(\theta)=\frac{1}{N_\mathrm{patch}}\sum_{n=1}^{N_\mathrm{patch}}\,w_{ij}^{(n)}\,(\theta)$.

We checked that the jackknife covariance matrix of each individual redshift bin is numerically stable with condition numbers ranging from $2.3\expo{4}$ for the covariance of the last redshift bin to $2.2\expo{5}$ for the first bin. Multiplying each of the jackknife covariances by its inverse yields the identity matrix, as expected from\,a\,numerically\,stable\,matrix,\,with\,a\,ratio\,between\,the diagonal and off-diagonal terms larger than $10^{13}$. In the Markov chain Monte Carlo (MCMC) analyses of Sect.~\ref{sc:Results}, the inverse of the covariance was corrected by the Hartlap multiplicative factor defined in \citet{Hartlap_2007}:
\begin{equation}
    \Psi = \frac{N_\mathrm{patch}-N_\mathrm{b}-2}{N_\mathrm{patch}-1},
\label{eq:hartlap}
\end{equation}
where $N_\mathrm{b}$ is the number of angular bins in the data vector.

The analytical covariance used in Sect.~\ref{sc:results_joint} is the sum of the Gaussian and super-sample contributions, leaving aside the connected non-Gaussian term from modes within the footprint. The Gaussian term was computed using \texttt{CosmoCov} as in \citet{Krause_2017} and \citet{Fang__2020}, with a correction for the survey footprint as in \citet{Troxel_2018} and without considering the Limber approximation \citep{Fang_2020}
\begin{multline}
\mathrm{Cov}^{\rm G}\left(C_{ij}(\ell),C_{kl}(\ell')\right) = 
\frac{4\,\pi\,\delta^{\rm K}_{\ell \ell'}}{\Omega_\mathrm{s} \,(2\ell+1)} \left[\left(C_{ik}(\ell)+ \frac{\delta^{\rm K}_{ik}}{\bar{n}_i}\right)\right. \\ \left.\times\left(C_{jl}(\ell')+\frac{\delta^{\rm K}_{jl}}{\bar{n}_j}\right) + \left(C_{il}(\ell) + \frac{\delta^{\rm K}_{il}}{\bar{n}_i}\right) \left(C_{jk}(\ell')+ \frac{\delta^{\rm K}_{jk}}{\bar{n}_j}\right)\right],
\end{multline}
with $C_{ij}$ the angular power spectrum of redshift bins $i$ and $j$, $\Omega_\mathrm{s}$ the survey area, $\bar{n}$ the effective number density of galaxies, and $i$, $j$, $k$, and $l$ are the indices of the redshift bins. The two-point angular correlation function was then computed from angular power spectra with \citepalias{Abbott_2024_DESY6}
\begin{multline}
\mathrm{Cov}^\mathrm{G}\left(w_{ij}\,(\theta),w_{kl}\,(\theta')\right)\\=\sum_{\ell,\ell'}\frac{(2\ell+1)\,(2\ell'+1)}{(4\,\pi)^2}\,\overline{\mathrm{P_\ell}}(\theta)\,\overline{\mathrm{P_{\ell'}}}(\theta')\,\mathrm{Cov}^\mathrm{G}\left(C_{ij}\,(\ell),C_{kl}\,(\ell')\right),
\label{eq:cov_2pcf}
\end{multline}
with $\overline{\mathrm{P_\ell}}$ the Legendre polynomial averaged over angular bins
\begin{equation}
\overline{\mathrm{P_\ell}} := \frac{\int_{x_\mathrm{min}}^{x_\mathrm{max}}\diff x\,P_\ell(x)}{x_\mathrm{max}-x_\mathrm{min}}=\frac{\left[P_{\ell+1}(x)-P_{\ell-1}(x)\right]_{x_\mathrm{min}}^{x_\mathrm{max}}}{(2\ell+1)(x_\mathrm{max}-x_\mathrm{min})},
\label{eq:legendre_poly_avg}
\end{equation}
in which $x=\cos(\theta)$ with $\thmin, \thmax$ the lower and upper limits of each angular bin.

As for the super-sample covariance (SSC) contribution to the covariance, it was computed following the fast approximation from \citet{Lacasa_2019} extended to partial-sky in \citet{Gouyou_Beauchamps_2022} to go beyond the full-sky approximation by taking into account the footprint of the survey
\begin{equation}
     \mathrm{Cov}^{\mathrm{SSC}}\left(w_{ij}\,(\theta),w_{kl}\,(\theta')\right) \approx \tilde{w}_{ij}\,(\theta) \ \tilde{w}_{kl}\,(\theta')\,S_{ijkl},
\label{eq:cov_ssc}
\end{equation}
where $\tilde{w}_{ij}\,(\theta)$ was computed as
\begin{equation}
    \tilde{w}_{ij}\,(\theta) = \sum_\ell \frac{2\ell+1}{4\pi} \ R_\ell \ C_{ij}(\ell) \ P_\ell(\cos\theta) =: (R \ast w_{ij})(\theta),
\end{equation}
with $R_\ell$ the response $\frac{\partial P_\mathrm{gal}(k,z)}{\partial\delta_\mathrm{b}}$ of the galaxy power spectrum to variations in the background density, $\delta_\mathrm{b}$. The $S_{ijkl}$ matrix element was computed using the implementation from \texttt{PySSC}\,\footnote{\url{https://github.com/fabienlacasa/PySSC}} as
\begin{equation}
S_{ijkl} = \int \diff V_1 \, \diff V_2 \, \frac{W_{i}(z_1)\,W_{j}(z_1)\,W_{k}(z_2)\, W_{l}(z_2)\,\sigma^2(z_1,z_2)}{\int \diff V_1 \ W_{i}(z_1) \ W_{j}(z_1)\,\int \diff V_2 \ W_{k}(z_2) \ W_{l}(z_2)},
\label{eq:sijkl}
\end{equation}
where $W_i$ is the kernel of the observable in redshift bin $i$, the kernel being the normalised redshift distribution $p_i(z)$ in the case of photometric galaxy clustering. The comoving volume element $\diff V = r^2(z)\,(\diff r/\diff z)\,\diff z$ was integrated between $z=0$ and $z=z_\mathrm{max}=3$. The variance of the background density $\sigma^2$ is, for a survey with a window function $\mathcal{W}$ (angular mask) of angular power spectrum $C^{\mathcal{W}}$:
\begin{equation}
\sigma^2(z_1,z_2) = \frac{1}{\Omega_\mathrm{s}^2}\sum_{\ell}(2\ell+1)\,C^{\mathcal{W}}(\ell)\,C(\ell,z_1,z_2),
\label{eq:sigma_ssc}
\end{equation}
where the angular matter power spectrum between redshifts $z_1$ and $z_2$ was obtained from the 3D matter cross-spectrum $P_\mathrm{m}(k\,|\,z_{12})=D\left(z_1\right)\,D\left(z_2\right)\,P_\mathrm{m}(k,z=0)$ with 
\begin{equation}
C(\ell,z_1,z_2) = \frac{2}{\pi}\,\int_{k_\mathrm{min}}^{k_\mathrm{max}}\,k^2\,\diff k\,P_\mathrm{m}(k\,|\,z_{12})\,j_\ell\left(kr_1\right)\,j_\ell\left(kr_2\right),
\label{eq:cl_z1_z2}
\end{equation}
where $D(z)$ is the linear growth factor, $r_i=r(z_i)$ is the comoving radial distance, $k_\mathrm{min}=0.1/r(z_\mathrm{max})$, and $k_\mathrm{max}=10/r(z_\mathrm{min})$.

We used the \texttt{anafast} routine from the \texttt{HEALPix}\footnote{\url{http://healpix.sf.net}} library \citep{Zonca2019,Gorski_2005} to compute $C^{\mathcal{W}}$. We made the approximation of a constant $R_\ell=5$, which reduces the convolution product to a multiplication. A detailed discussion on the response of the SSC can be found in \citet{Sciotti_2024}. The approximation on $R_{\ell}$ used in this work has an impact that does not exceed 0.5\% on $\alpha$ and 3\% on its uncertainty in all redshift bins, which is expected given the angular scales and redshifts considered.

\section{Methodology}
\label{sc:methodology}
\subsection{Galaxy bias}
\label{sc:bg}
A fit of the linear galaxy bias $b(z)$ was performed in each redshift bin using the jackknife covariance and the residuals $w(\theta) - b^2w_{\mathrm{fid,}b=1}(\theta)$, where $w_{\mathrm{fid,}b=1}$ denotes the theoretical angular correlation function defined in Eq.~\eqref{eq:w_theory} computed with the Flagship cosmology and a galaxy bias $b=1$. We used scales between $0\degf5$ and $4^\circ$ in this full shape fit. A third-order polynomial was then fitted to the result
\begin{equation}
b(z) = b_3 z^3 + b_2 z^2 + b_1 z + b_0,
\label{eq:bg}
\end{equation}
which was then used in Eq.~\eqref{density_field_theory} to compute the theoretical model of the two-point angular correlation function.

\subsection{BAO template fitting}
\label{sc:methodology_BAO}
To extract the BAO feature from the photometric sample, we performed the fitting between the measured $w(\theta)$ and a template derived from the theoretical two-point angular correlation function $w_\mathrm{fid}(\theta)$ computed for the fiducial cosmology described in Sect.~\ref{sc:Data}. The template is defined as
\begin{equation}
T\left(\alpha,\theta\right) := B\,w_{\rm fid}\left(\alpha \theta\right)+A_0+\frac{A_1}{\theta}+\frac{A_2}{\theta^2},
\label{eq:template_BAO}
\end{equation}
where $\alpha$ is the transverse Alcock--Paczynski parameter quantifying an eventual shift of the BAO peak between the measured $w(\theta)$ and the fiducial $w_\mathrm{fid}(\theta)$. The nuisance parameters $B$, $A_0$, $A_1$, and $A_2$ are needed to marginalise over the shape of $w(\theta)$ and absorb residual effects such as non-linear galaxy bias to only extract information from the BAO peak. Different template parametrisations can be used and we verify in Sect.~\ref{sc:results_templates} that the choice of polynomial has minimal impact on the $\alpha$ parameter. The parameter of interest in this work is $\alpha$ and, since the fiducial cosmology used to compute $w_\mathrm{fid}(\theta)$ is the same as the simulation, we expect to recover $\alpha=1$. On the contrary, using a different fiducial cosmology to compute $w_\mathrm{fid}(\theta)$ should result in $\alpha \neq 1$.

We considered one set of nuisance parameters per redshift bin, since these parameters can in principle vary with redshift. This represents a total of 53 parameters for the joint analysis and 5 parameters for the analysis of each of the 13 bins. In the joint analysis, we included the cross-covariance from the redshift bins. The MCMC technique was used to quantify the uncertainty on $\alpha$ marginalised over the nuisance parameters. The \texttt{emcee} sampler introduced in \citet{Foreman_Mackey_2013} was used with the Gelman--Rubin convergence stopping criterion described in \citet{Gelman_1992} with a threshold $R_\mathrm{GR}-1 = 0.005$ for analyses on individual redshift bins and $R_\mathrm{GR}-1=0.02$ for joint analyses. These thresholds were chosen to stop chains when parameter values and uncertainties reached a plateau. Uniform priors applied to the template parameters are presented in Table~\ref{tab:priors}.

\begin{table}
\centering
\caption{Priors used for the template-fitting parameters.}
\begin{tabular}{cccc}
\hline
\hline
\noalign{\vskip 2pt}
$\alpha$ & $10^3\,A_0$ & $10^3\,A_i$ ($i\neq0$) & $B$\\
\hline
\noalign{\vskip 2pt}
[$0.8$,$1.2$] & [$-1$,$1$] & [$-5$,$5$] & [$0.2$,$6$]\\
\hline
\end{tabular}
\tablefoot{With the first template, defined in Eq.~\eqref{eq:template_BAO}, the $A_i$ parameters have a unit of $\mathrm{deg}^{i}$. For the other templates, defined in Sect.~\ref{sc:results_templates}, the units should be adapted to have a template $T(\alpha,\theta)$ of unit consistent with $w(\theta)$.}
\label{tab:priors}
\end{table}

As a comparison to MCMC, we also considered the frequentist approach of profile likelihood to provide constraints in the joint analysis. We obtained the profile likelihood 
$\Delta\chi^2(\alpha) = \chi^2(\alpha) - \left.\chi^2(\alpha)\right|_\mathrm{min}$ by computing the best fit $\chi^2$ for each value of $\alpha$ between 0.8 and 1.2 in steps of 0.001. We then fitted an 8th-order polynomial to this profile to compute $\Delta\chi^2$. The $1\,\sigma$ uncertainty on $\alpha$ is then given by $\Delta\chi^2=1$. In the joint analysis, the resolution of the grid of $\alpha$ on which this polynomial was evaluated was increased to use steps of 0.0001 to match the increased constraining power. We also used this frequentist approach to quantify the significance of the BAO detection as
\begin{equation}
\Delta_\mathrm{det} := \left|\chi^2_\mathrm{w}(\alpha_\mathrm{min})-\chi^2_\mathrm{no\,wiggle}(\alpha_\mathrm{min})\right|^{1/2},
\label{eq:bao_detection_level}
\end{equation}
evaluated at the $\alpha_\mathrm{min}$ value minimising $\chi^2_\mathrm{w}(\alpha)$. The $\chi^2_\mathrm{no\,wiggle}$ was computed using the $w(\theta)$ in which the BAO peak was removed with the filter proposed in \citet{Kirkby_2013}, unlike the $w(\theta)$ used to obtain $\chi^2_\mathrm{w}$. The code \texttt{iminuit} based on the MINUIT algorithm was used to this effect \citep{iminuit,James:1975dr}.

\subsection{Cosmological parameters}
\label{sc:methodology_cosmology}

Extracting the transverse Alcock--Paczynski $\alpha$ parameter in successive tomographic bins of redshift allows us to constrain the evolution of the expansion of the Universe. Indeed, the $\alpha$ parameter can be expressed as
\begin{equation}
\alpha = \frac{D_\mathrm{A}}{r_\mathrm{d}}\;\frac{r_\mathrm{d,\,fid}}{D_\mathrm{A,\,fid}},
\label{eq:alpha}
\end{equation}
where $D_\mathrm{A}$ is the angular diameter distance, $r_\mathrm{d}$ corresponds to the sound horizon at the drag epoch, and the fid label stands for the values in the fiducial cosmology. The details of the computation of $r_\mathrm{d}$ can be found in Appendix~\ref{ap:rdrag}. We sampled over $h\,r_\mathrm{d}$ and $\Omm$ with MCMC, comparing the $\alpha_i$ value for each redshift bin to the theoretical expected value of the fiducial cosmology. In Eq.~\eqref{eq:alpha}, the respective quantities are all evaluated at the effective redshift. The effective redshift of bin $i$ is defined as
\begin{equation}
z_{\mathrm{eff},i} := \int_0^{z_\mathrm{max}}\,\diff z\,z\,p_i(z),
\label{eq:effective_redshift}
\end{equation}
where $p_i(z)$ is the normalised distribution of photometric redshifts shown in Fig.~\ref{fig:flagship_nz}. We checked that using the median redshift of each redshift bin did not affect the constraints on $h\,r_\mathrm{d}$ and $\Omm$ significantly (shifts of $0.06\,\sigma$ and $0.02\,\sigma$, respectively). We used large uniform priors $\Omm\in\mathcal{U}\left[0.1, 0.9\right]$ and $h\,r_\mathrm{d}\in\mathcal{U}\left[20, 1000\right]\,\mathrm{Mpc}$.

\begin{figure*}
    \vspace{-2pt}
    \centering
    \includegraphics[width=\textwidth]{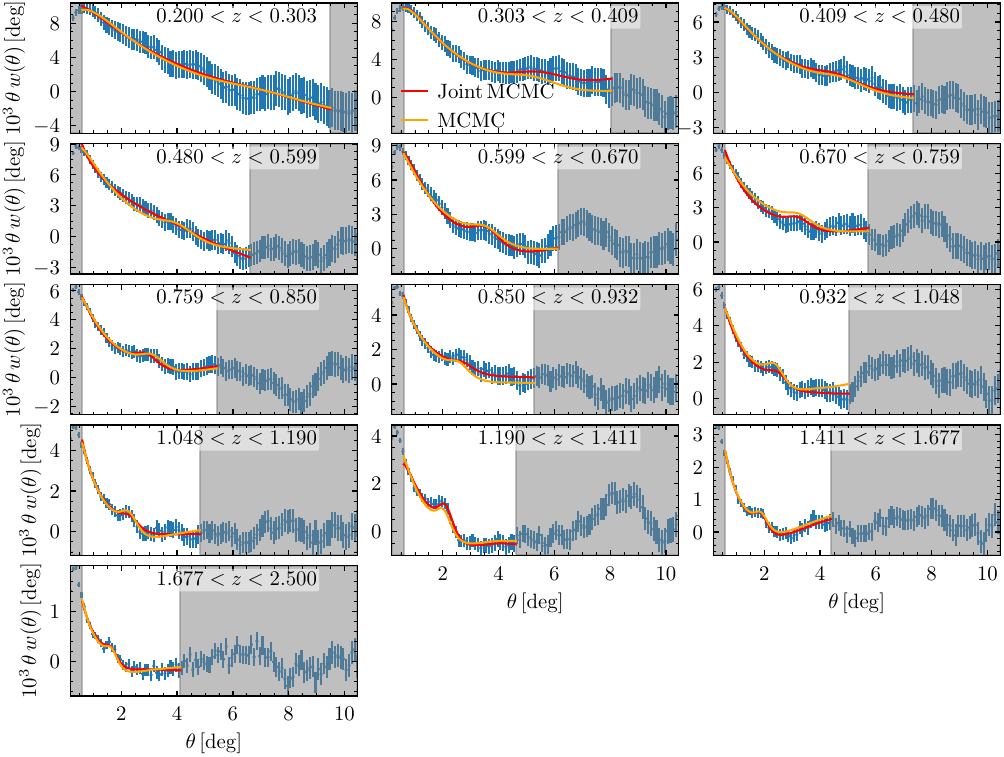}
    \caption{Two-point angular correlation function measured on the Flagship simulation in 13 redshift bins. The errors come from the analytical covariance presented in Sect.~\ref{sc:covariance}. The orange curve is the correlation function computed using the template from Eq.~\eqref{eq:template_BAO} evaluated with the parameters inferred from MCMC in each redshift bin while the red curve uses parameters from a joint analysis of all redshift bins. Scale cuts are shown as grey bands and are defined as $\thmin=0\degf6$, $\thmax=\thbao+2\degf5$, where $\thbao$ is the expected position of the BAO peak in the fiducial cosmology.}
    \vspace{-4pt}
    \label{fig:flagship_2pcf}
\end{figure*}

\section{Results}
\label{sc:Results}

\subsection{Individual bins BAO measurement}
\label{sc:results_individual_bins}

In \,this \,section, \,we \,present \,the \,constraints obtained by applying the template fit to one redshift bin at a time, yielding 13 values of the $\alpha$ parameter. This reduces the constraining power over each $\alpha$ with respect to a joint analysis but gives information about the redshift evolution that can be used to constrain cosmological parameters as explained in Sect.~\ref{sc:methodology_cosmology}. It is also a more relevant analysis in our setup given the caveat of fitting a unique $\alpha$ explained at the end of Sect.~\ref{sc:results_joint}. The scale cuts used are $\thmin=0\degf6$, $\thmax=\thbao+2\degf5$, visible as grey bands in Fig.~\ref{fig:flagship_2pcf} and are discussed in detail in Sect.~\ref{sc:results_cuts}. We clearly observe that the BAO peak shifts to smaller angular separations as redshift increases, moving from $7^\circ$ in the first redshift bin to $1\degf6$ in the last. We study the impact of a different choice of scale cuts in Sect.~\ref{sc:results_cuts}.

\begin{figure}
    \vspace{3pt}
    \centering
    \resizebox{\hsize}{!}{\includegraphics{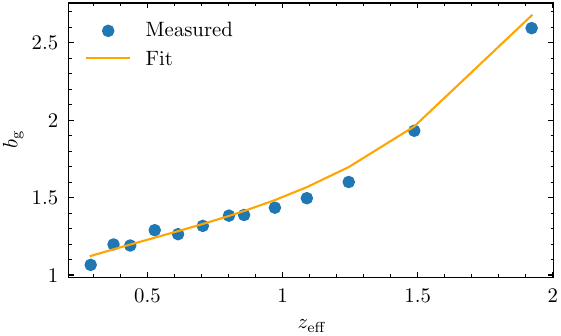}}
    \caption{Galaxy bias measured on Flagship 2.1 between $0\degf5$ and $4^\circ$ and its polynomial fit.}
    \label{fig:bg_fit}
\end{figure}

We first report the estimate of the linear galaxy bias, shown in Fig.~\ref{fig:bg_fit} with the best fit obtained for $b_3=0.2681$, $b_2=-0.4090$, $b_1=0.6944$, and $b_0=0.9493$ as the coefficients of Eq.~\eqref{eq:bg}. These values of the bias coefficients are then used to compute the fiducial $w_{\rm fid}(\theta)$ that is injected in the template of Eq.~\eqref{eq:template_BAO}. The shape of $w(\theta)$ and eventual non-linearities of the galaxy bias at small scales or systematic effects are absorbed by the template nuisance parameters without affecting $\alpha$, our parameter of interest.

\begin{figure}
    \centering
    \includegraphics[width=0.47\textwidth]{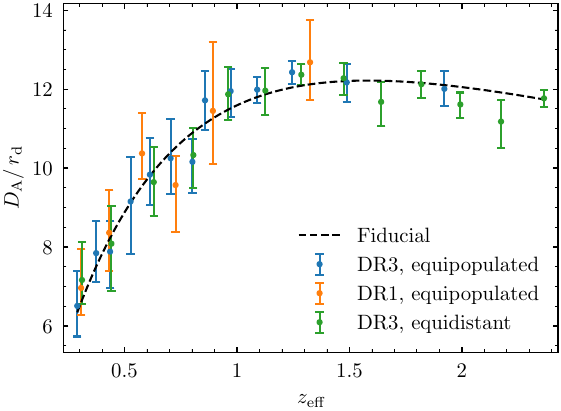}
    \caption{Comparison of the constraints on $D_\mathrm{A}/r_\mathrm{d}$ with equipopulated and equidistant redshift bins for DR3 and DR1.}
    \label{fig:da_rd}
\end{figure}

Table~\ref{tab:alpha_i} groups the results for $\alpha$ in all 13 redshift bins, along with the associated sigma level of BAO detection. We first notice that the $1\,\sigma$ uncertainty is larger at low redshift. This is due to the smearing of the BAO signal by the variations in the peak position in each bin, in addition to the non-linear evolution of the large-scale structures under the effect of gravitation. This is clearly visible in Fig.~\ref{fig:flagship_2pcf} where the BAO signal is much more peaked at higher redshift. The parameter $\alpha$ is compatible with $\alpha=1$ within $1\,\sigma$ in all redshift bins with the exception of bin 11 at $z_\mathrm{eff}=1.245$ for which we find a $1.3\,\sigma$ shift. This bin is also the one with the strongest constraining power, explained by the high redshift and small scatter of photometric redshifts $\sigma_\mathrm{NMAD}=0.024$. As for the level of detection of the BAO signal, we find four redshift bins with no significant detection, bins 1, 3, 4, and 6. Otherwise, the significance of the detections ranges between 1.1 and $4.0\,\sigma$ detections, with a maximum at $z_\mathrm{eff}=1.245$ and the $1\,\sigma$ uncertainty on $\alpha$ decreases as the detection level increases, as expected. The evolution of $D_\mathrm{A}/r_\mathrm{d}$ as a function of redshift can be seen in Fig.~\ref{fig:da_rd} along with the results obtained with a DR1-like selection and a different binning scheme, discussed, respectively, at the end of this section and in Sect.~\ref{sc:results_equidist_binning}.

\begin{table}
\centering
\caption{Values of $\alpha$ extracted from the MCMC analysis in each of the 13 redshift bins.}
\setcellgapes{1pt}\makegapedcells
\begin{tabular}{lcccc}
\hline
\hline
\noalign{\vskip 5pt}
 Bin & $z_\mathrm{eff}$ & $\alpha$ & $D_\mathrm{A}/r_\mathrm{d}$ & $\Delta_\mathrm{det}$ ($\sigma$) \\
     \noalign{\vskip 1pt}
  \hline
  \noalign{\vskip 2pt}
      1 & 0.290 & $1.026^{+0.122}_{-0.140}$ & $6.509^{+0.774}_{-0.888}$ & --  \\
      2 & 0.374 & $1.044^{+0.097}_{-0.107}$ & $7.847^{+0.729}_{-0.804}$ & 1.1     \\
      3 & 0.436 & $0.957^{+0.112}_{-0.093}$ & $7.885^{+0.923}_{-0.766}$ & --     \\
      4 & 0.527 & $1.003^{+0.146}_{-0.123}$ & $9.155^{+1.333}_{-1.123}$ & --  \\
      5 & 0.613 & $1.002^{+0.079}_{-0.095}$ & $9.835^{+0.775}_{-0.932}$ & 1.1    \\
      6 & 0.705 & $0.985^{+0.087}_{-0.096}$ & $10.249^{+0.905}_{-0.999}$ & --  \\
      7 & 0.802 & $0.932^{+0.072}_{-0.054}$ & $10.160^{+0.785}_{-0.589}$ & 1.8    \\
      8 & 0.858 & $1.052^{+0.067}_{-0.067}$ & $11.716^{+0.746}_{-0.746}$ & 1.8    \\
      9 & 0.972 & $1.037^{+0.057}_{-0.048}$ & $11.948^{+0.657}_{-0.553}$ & 2.3    \\
      10 & 1.090 & $1.015^{+0.029}_{-0.028}$ & $11.988^{+0.343}_{-0.331}$ & 3.0    \\
      11 & 1.245 & $1.031^{+0.024}_{-0.024}$ & $12.429^{+0.289}_{-0.289}$ & 4.1    \\
      12 & 1.488 & $0.996^{+0.040}_{-0.038}$ & $12.171^{+0.489}_{-0.464}$ & 2.5    \\
      13 & 1.922 & $0.991^{+0.036}_{-0.037}$ & $12.009^{+0.436}_{-0.448}$ & 2.9    \\
\hline
\end{tabular}
\tablefoot{The detection level is defined in Eq.~\eqref{eq:bao_detection_level}. When the significance is smaller than $1\,\sigma$, the result is considered as a non-detection and indicated with a dash.}
\label{tab:alpha_i}
\end{table}

When averaged over all redshift bins, the shift between the results obtained with the jackknife and with the analytical covariances is smaller than $0.3\,\sigma$ with the measured data vector and $0.08\,\sigma$ with a noise-free synthetic data vector computed similarly to the theoretical model, used to check that the shifts observed with respect to $\alpha=1$ are explained by the quality of the measured $w(\theta)$.

\begin{figure}
    \centering
    \resizebox{\hsize}{!}{\includegraphics{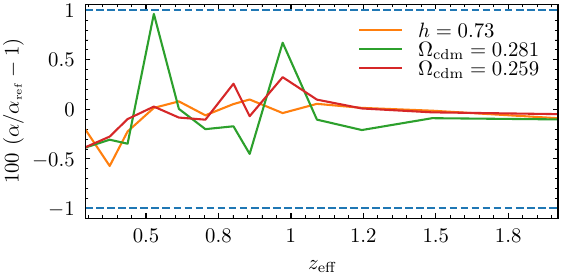}}
    \caption{Effect of the fiducial cosmology on $\alpha$ in all redshift bins for three variations replacing $h=0.67$ by $h=0.73$ and $\Omcdm=0.270$ by $\Omcdm=0.281$ or $\Omcdm=0.259$. The dashed line shows a relative difference of 1\%. The effect of the fiducial cosmology is negligible with respect to the uncertainties varying between $\pm0.13$ and $\pm0.024$.}
    \label{fig:alpha_cosmo_fid}
\end{figure}

The impact of the choice of fiducial cosmology was investigated by varying $h$ from 0.67 to 0.73. Galaxy bias was fitted again before repeating the MCMC analysis bin by bin. We expect a shift of $\alpha$ by a factor 
\begin{equation}
\frac{D_\mathrm{A}(h=0.73, \Omm)}{r_\mathrm{d}(h=0.73, \omb, \omm)}\;\frac{r_\mathrm{d}(h=0.67, \omb, \omm)}{D_\mathrm{A}(h=0.67, \Omm)} = 0.982.
\end{equation}
After correction by this shift, we measure a remaining maximum relative difference $|\Delta\alpha|/\alpha$ of 0.01, the average over all redshift bins being 0.0012. This is illustrated in Fig.~\ref{fig:alpha_cosmo_fid}. As an additional test, we varied $\Omcdm$ by $\pm5\,\sigma_{Planck}$, keeping $h=0.67$ and $\Omb=0.049$. With $\Omcdm=0.281$ and $\Omcdm=0.259$, we find a remaining maximum relative difference $|\Delta\alpha|/\alpha$ of, respectively, 0.9\% and 0.4\%, the average over all redshift bins being 0.31\% and 0.13\%. These variations are negligible compared to the uncertainties on $\alpha$, which shows that the analysis is robust against the choice of fiducial cosmology.

\begin{table}
\centering
\caption{Values of $\alpha$ extracted from the MCMC analysis in a DR1-like setting in each equipopulated redshift bin.}
\setcellgapes{1pt}\makegapedcells
\begin{tabular}{lcccc}
\hline
\hline
\noalign{\vskip 5pt}
 Bin & $z_\mathrm{eff}$ & $\alpha$ & $D_\mathrm{A}/r_\mathrm{d}$ & $\Delta_\mathrm{det}\,(\sigma)$\\
\noalign{\vskip 1pt}
\hline
\noalign{\vskip 2pt}
1 & 0.307 & $1.055^{+0.102}_{-0.148}$ & $6.964^{+0.673}_{-0.977}$ & -- \\
 2 & 0.432 & $1.021^{+0.118}_{-0.131}$ & $8.364^{+0.967}_{-1.073}$ &  --  \\
 3 & 0.578 & $1.086^{+0.068}_{-0.106}$ & $10.372^{+0.649}_{-1.012}$ & 1.3  \\
 4 & 0.727 & $0.909^{+0.113}_{-0.070}$ & $9.571^{+1.190}_{-0.737}$ & 1.1   \\
 5 & 0.893 & $1.016^{+0.120}_{-0.155}$ & $11.450^{+1.352}_{-1.747}$ &  --  \\
 6 & 1.325 & $1.045^{+0.079}_{-0.089}$ & $12.680^{+0.959}_{-1.080}$ & 1.1 \\
\hline
\end{tabular}
\label{tab:alpha_i_dr1}
\end{table}

We also provide constraints in a DR1-like setting with a sample divided into six redshift bins. The measurement of $w(\theta)$ and BAO analysis were done following the same procedure. In this case, we find that the constraints on $\alpha$ in bins 1 to 6, listed in Table~\ref{tab:alpha_i_dr1}, are in agreement with $\alpha=1$ within $1\,\sigma$. If we compare bins of similar effective redshifts, the constraints are about 20\% weaker than with 13 bins in the first bins and significantly worse in the last two bins. The detection of the BAO signal is overall weaker than with 13 redshift bins, with no significant detection in bins 1, 2, and 5. These results are expected from the larger uncertainties on $w(\theta)$ and the larger bins: intra-bin variations in the BAO scale dilute the signal. Note that the LSST-like photometry assumed to infer photometric redshifts is even more optimistic for this setting than for the previous one, since this photometry will not be available at the time of this data release. Instead, photometry from the Dark Energy Survey will be used \citep{DES_DR1}. For this reason, the fact that the redshift distribution $p(z)$ is well known is only true with the Flagship simulation. With data, calibrating the $p(z)$ bias and stretch prior to the analysis will be mandatory, as in \citetalias{Abbott_2024_DESY6}. Ideally, these nuisance parameters for the bias and stretch of $p(z)$ will be marginalised over in the MCMC analysis of DR1 data as in \citet{Bertmann_2026}. These constraints could probably be improved with analysis choices tailored to this sample, for example with different scale cuts.

 \subsection{Cosmological constraints from BAO}
 \label{sc:results_individual}

\begin{figure}
    \centering
    \resizebox{\hsize}{!}{\includegraphics{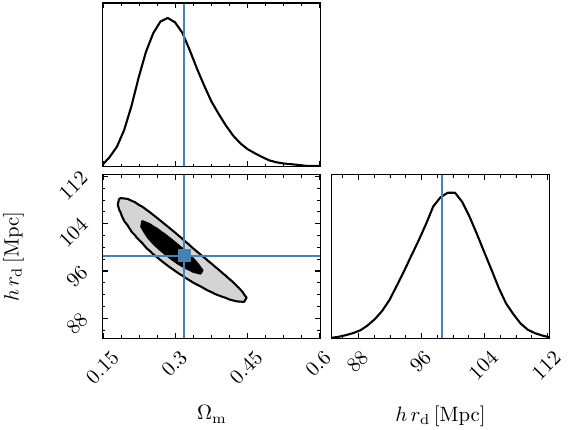}}
    \caption{Constraints on $h\,r_\mathrm{d}$ and \Omm\, from BAO. They are in agreement with the fiducial values $h\,r_\mathrm{d}=98.58$ Mpc and $\Omm=0.319$, shown in blue, within $0.34\,\sigma$ and $0.18\,\sigma$, respectively.}
    \label{fig:corner_plot_alpha_fit}
\end{figure}

 One can constrain \Omm\, and $h\,r_\mathrm{d}$ by sampling them in their dependence with respect to the values of $\alpha_{i}$, $i\in[1,13]$ (Eq.~\ref{eq:alpha} and Appendix~\ref{ap:rdrag}) obtained by template fitting in each individual bin. We list in Table~\ref{tab:alpha_i} the values of $\alpha_{i}$ and the associated uncertainties obtained by MCMC analysis. In order to account for correlations between different redshift bins, we estimate their covariance from the $\alpha_{i}$ values extracted from $w(\theta)$ measured in the jackknife samples. The first three bins are excluded from the cosmological analysis because the model itself contains too little BAO signal there: this is confirmed by obtaining $\Delta_\mathrm{det} < 1\,\sigma$ when fitting a noise-free synthetic data vector instead of the measured $w(\theta)$. In Fig.~\ref{fig:corner_plot_alpha_fit}, we show the constraints obtained on $h\,r_\mathrm{d}$ and \Omm. We measure $\Omm=0.296^{+0.074}_{-0.059}$ and $h\,r_\mathrm{d}=99.35^{+4.29}_{-4.55}\,\mathrm{Mpc}$, in agreement with the simulation cosmology within $0.34\,\sigma$ and $0.18\,\sigma$, respectively. We acknowledge that these constraints are weak when compared to DESI DR2, which yields $\Omm=0.2975 \pm 0.0086$, $h\,r_\mathrm{d}=101.54 \pm 0.73\,\mathrm{Mpc}$ \citep{Abdul_Karim_2025}. This is explained by the fact that we only use the transverse Alcock--Paczynski parameter and not the radial component $\alpha_\parallel$ and by the lack of optimisation of the galaxy sample and redshift binning.

\begin{figure}
    \centering
    \resizebox{\hsize}{!}{\includegraphics{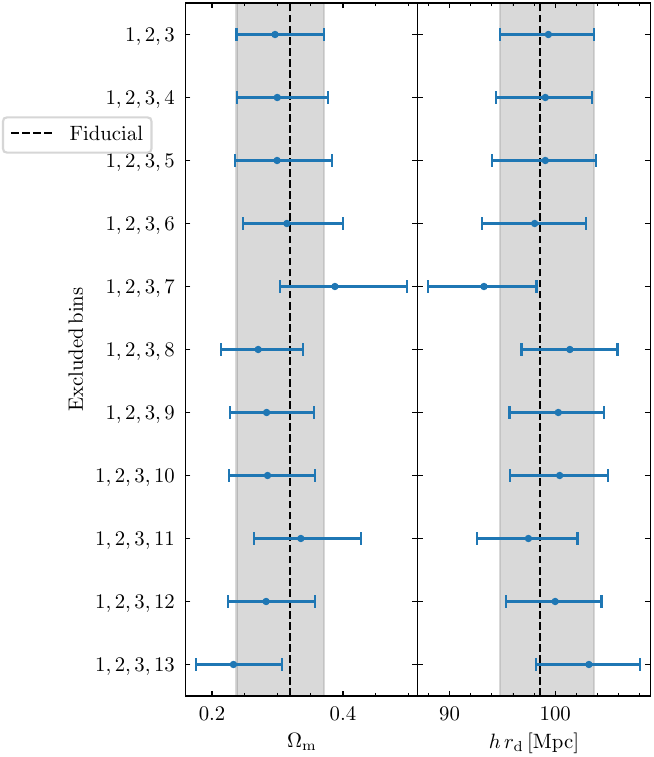}}
    \caption{Comparison of cosmological parameters obtained when redshift bins are removed from the analysis. The fiducial values of the simulation are highlighted as a vertical dashed line. The baseline including bins 4 to 13 is at the top as a reference, with its associated uncertainty shown in grey.}
    \label{fig:cosmo_bins_robustness}
\end{figure}

We checked how excluding different redshift bins from this analysis affects the cosmological constraints. The baseline analysis excludes the first three redshift bins so we consider the $\alpha_{i}$, $i\in[3,13]$ obtained by template fitting on $w(\theta)$ measured on Flagship. The associated result is shown at the top of Fig.~\ref{fig:cosmo_bins_robustness} for \Omm\, (left panel) and $h\,r_\mathrm{d}$ (right panel), with the uncertainty of the baseline in grey. Then, we removed one redshift bin at a time. Redshift bins 3 to 6 have only a marginal effect on the value of \Omm\, and $h\,r_\mathrm{d}$, with a slight increase in the uncertainty of 4\% to 15\% for \Omm\, and 3\% to 11\% for $h\,r_\mathrm{d}$. Redshift bin 7 seems to have a large weight in shifting \Omm\, and $h\,r_\mathrm{d}$ towards smaller and larger values, respectively: when it is excluded, we find estimates of \Omm\, and $h\,r_\mathrm{d}$ that are shifted by 1 and $1.2\,\sigma$ with respect to the fiducial values, in units of the baseline uncertainty, while the uncertainties are 46\% and 16\% larger. Removing the other bins has a much smaller effect, with variations of the uncertainties that are within 7\%. Two exceptions are bin 11, which increases the uncertainty of \Omm\, by 22\% owing to its strong constraining power, and bin 13, which shifts \Omm\, by $1.3\,\sigma$ and $h\,r_\mathrm{d}$ by $1\,\sigma$ while increasing the uncertainty of the latter by 11\%.
\subsection{Joint BAO measurement}
\label{sc:results_joint}
We now present the constraints on $\alpha$ obtained with a joint analysis of the 13 redshift bins. We used the template as defined in Sect.~\ref{sc:methodology_BAO} extended to have bin-specific nuisance parameters $B_i$, $A_{0,i}$, $A_{1,i}$, and $A_{2,i}$ with $i\in[1,13]$. We used the analytical covariance (Gaussian and SSC) for this joint analysis.

For this joint analysis with all redshift bins, we report a constraint of $\alpha = 1.0011^{+0.0078}_{-0.0079}$, obtained with a profile likelihood approach \citep{Planck_2014}. This approach consists in minimising the $\chi^2$ while fixing a parameter at various values to obtain a profile of $\chi^2$ as a function of this parameter. Repeating this with another model of $w(\theta)$, we can then obtain the difference of the profiles, $\Delta\chi^2$. This result might seem optimistic given that it represents an improvement by a factor of 3 with respect to the latest results from the \citetalias{Abbott_2024_DESY6} analysis, which yields a constraint of $\alpha$ at the 2.4\% level with the angular two-point correlation function. However, several factors need to be taken into account in this comparison. The first one is that this work uses data from a simulation and so it is inherently optimistic since it is free from systematic effects while \citetalias{Abbott_2024_DESY6} uses real observations and has to correct them. 
In \citetalias{Abbott_2024_DESY6}, 5 redshift bins between $0.7 < z < 1.2$ are used whereas the Flagship sample is here divided into 13 redshift bins between $0.2 < z < 2.5$, which increases significantly the constraining power of the joint analysis. We quantified the significance of the BAO detection defined in Sect.~\ref{sc:methodology_BAO} to be $\Delta_\mathrm{det}=10.3\,\sigma$ with our data compared to $3.5\,\sigma$ in \citetalias{Abbott_2024_DESY6}. We also find significantly tighter constraints in each individual redshift bin, contributing to this result. 
\begin{figure}
    \centering
    \resizebox{\hsize}{!}{\includegraphics{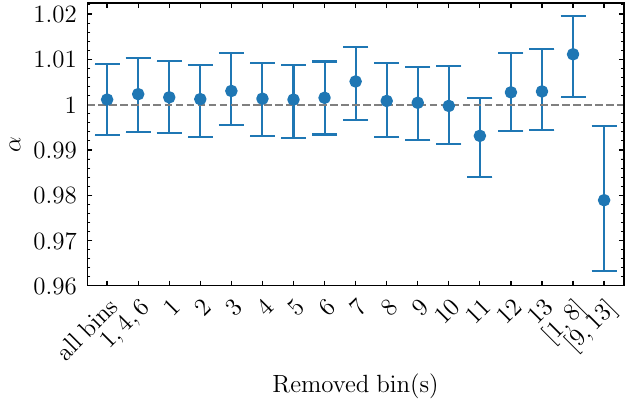}}
    \caption{Comparison of $\alpha$ from a joint analysis in which redshift bins are successively removed. Bin 11 with effective redshift $z_\mathrm{eff}=1.245$ has a larger effect on the joint constraints compared to other bins, as one can see from the shift and increase in the uncertainty when it is not included. Selecting high- or low-redshift bins, respectively, induces a 1.3 or $2.8\,\sigma_\mathrm{all\;bins}$ shift towards higher or lower values for $\alpha$.}
    \label{fig:bins_robustness}
\end{figure}
For this joint analysis, we find that the results obtained with profile likelihood and MCMC differ significantly because of the combined effect of the strong constraining power and the interplay between the theoretical model and our data at scales smaller than $1^\circ$, which creates oscillations in the posterior distribution. This results in a poor exploration of a noisy posterior distribution where the MCMC becomes trapped in a local maximum, whereas the polynomial fit used in the profile likelihood methodology introduced at the end of Sect.~\ref{sc:methodology_BAO} circumvents this issue. Investigations of this issue are provided in Appendix~\ref{ap:proflikecomp}. The poor exploration by the MCMC persists even when we increased the standard deviation of the initialisation values of the walkers from $10^{-4}$ to $10^{-2}$, when we changed the move strategy or adopted a larger stretch scale parameter. Increasing the Core Cosmology Library internal spline parameter $\mathrm{ELL\_MAX\_CORR}$ controlling the maximum $\ell$ used in the FFTLog computation of $w(\theta)$ from its default value of 60\,000 to 150\,000 completely removes these oscillations. We then obtain a joint constraint of $\alpha=1.0055^{+0.0081}_{-0.0080}$. The constraint obtained from MCMC with this new parameter value is $\alpha = 1.0049 \pm 0.0059$, in better agreement with the profile likelihood result.

We further investigate this result by using a similar approach to \citetalias{Abbott_2024_DESY6} by excluding redshift bins in which the BAO signal is not detected with sufficient strength. In this case, $\Delta_\mathrm{det}$ is computed for each individual redshift bin and a non-detection is then defined as a detection level $\Delta_\mathrm{det} < 1$. A non-detection in the first bin ($0.6 < z < 0.7$) is the reason why 5 bins were used in the analysis of \citetalias{Abbott_2024_DESY6} instead of 6. 
After excluding the redshift bins with no significant detection, we report a value $\alpha=1.0023^{+0.0080}_{-0.0084}$, almost identical to the result obtained when including all the redshift bins. The fact that the increase in the uncertainty is as small as 4.5\% despite removing three out of 13 bins can be understood by the fact that if redshift bins have no significant detection of the BAO signal then they provide very little constraining power on $\alpha$. We check the robustness of our result with respect to all the redshift bins included in the joint analysis in Sect.~\ref{sc:results_robustness}.

An \,important \,caveat \,to \,fitting \,a \,unique \,$\alpha$ \,to \,all  redshift \,bins \,is \,that \,it \,assumes \,a \,perfect \,match \,between \,the \,fiducial \,cosmology \,and \,the \,true \,cosmology \,of \,the \,Universe, \,which \,is unknown. Any mismatch leads to a variation in $\alpha$ that depends \,on \,redshift. \,This \,effect \,can \,be \,quantified \,thanks \,to \,the \,definition of $\alpha$ in Eq.~\eqref{eq:alpha} in which $D_\mathrm{A}$ and $r_\mathrm{d}$ are computed for different cosmologies while $D_\mathrm{A,\,fid}$ and $r_\mathrm{d,\,fid}$ are constant at the fiducial cosmology of the Flagship simulation. We checked that varying $\Omcdm$, $\Omb$, and $h$ by 5\% leads to a maximum expected variation in $\alpha$ of 1\% between the first and last redshift bin. This maximum variation in $\alpha$ scales linearly as $\alpha(z_\mathrm{eff}=1.922)/\alpha(z_\mathrm{eff}=0.290)\propto0.2\,\Delta(\Omcdm, \Omb,h)$. This corresponds to having a one sigma shift between the first and last redshift bin (joint constraint of 0.78\%) if the fiducial cosmology is wrong with respect to the true one by $2.1\,\sigma_{Planck18}(\Omcdm)$ and $2.3\,\sigma_{Planck18}(\Omb)$ at the same time \citep[TT,TE,EE+lowE+lensing]{Aghanim_2018}. Given the constraint obtained on $\alpha$, this effect is non-negligible and is a limit to this joint analysis. For this reason, the main analysis is performed in individual redshift bins in Sect.~\ref{sc:results_individual_bins}.
\vspace{0.03cm}
\subsection{Robustness validation}
\label{sc:results_robustness}

In this section, the impact of the different redshift bins in the joint analysis is evaluated by removing each of the 13 redshift bins, one at a time. The constraints on $\alpha$ for all these cases are shown in Fig.~\ref{fig:bins_robustness}. While most bins have very little effect on the constraints, removing the redshift bin 11 at $z_\mathrm{eff}=1.245$ from the joint analysis decreases $\alpha$ by $1\,\sigma_\mathrm{all\;bins}$. It also increases uncertainties by 11\%, which is expected given that we remove one of the redshift bins with the most constraining power. This constraining power can be understood by studying the properties of the photometric redshifts computed for the Flagship galaxy mock catalogue. Indeed, comparing them to the true redshifts of the simulation using the same binning, we find that with a measure of the scatter $\sigma_\mathrm{NMAD}$ robust to outliers and defined in Eq.~\eqref{eq:sigma_nmad}, bin 11 has a scatter that is about 25\% smaller than bins 12 and 13. The combination of good photometric redshifts and high redshift explains the importance of this bin.

If we compare constraints from bins with redshift $z_\mathrm{eff}<0.9$ to the baseline with all bins, we find that $\alpha$ is shifted by $2.8\,\sigma_\mathrm{all\;bins}$ towards smaller values and its uncertainty is increased by a factor of two. On the contrary, including high-redshift bins $z_\mathrm{eff}>0.9$ and removing low-redshift bins, the shift towards a larger value of $\alpha$ is limited to $1.3\,\sigma_\mathrm{all\;bins}$. The uncertainty is only increased by 14.6\%, in this case. These results can be understood in light of the constraints from individual bins detailed in Table~\ref{tab:alpha_i}. Bins at $z_\mathrm{eff}<0.9$ with the largest level of significance of BAO detection are biased towards low values of $\alpha_i$, which explains why the joint value of $\alpha$ increases when they are removed. On the other hand, bins at $z_\mathrm{eff}>0.9$ are overall biased towards larger values of $\alpha_i$, which explains why removing them decreases the value of the joint $\alpha$. The large difference in the uncertainty values for these last two cases ($1.3$ against $2.8\,\sigma_\mathrm{all\;bins}$) can be explained by the much tighter constraints obtained at high redshift, where the BAO peak is not smeared.
\vspace{9pt}
\subsection{Comparison of fitting templates}
\label{sc:results_templates}
 \begin{figure*}
    \vspace{3pt}
    \centering
    \includegraphics[width=\textwidth]{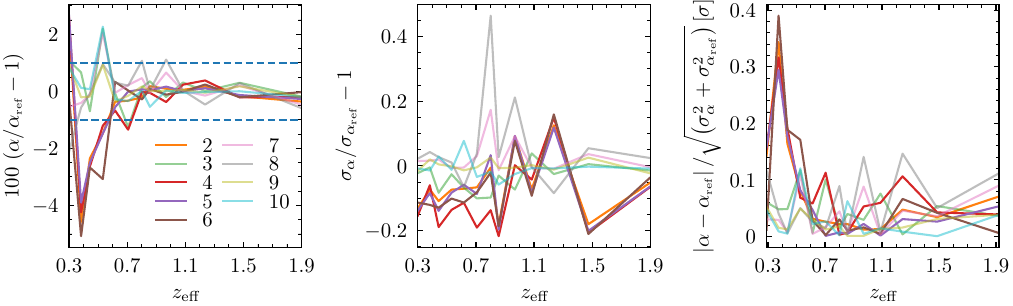}
    \caption{Comparison of templates for BAO fitting as a function of the redshift bin. The first panel represents the shift of $\alpha$ with respect to the reference (template 1). We highlight by transparency templates 2, 4, 5, and 6, which do not include a term in $\theta^{-2}$ to showcase the fact that they all have larger shifts at low redshift. The 1\% dashed blue lines are a reference to guide the eye rather than a goal, given that the uncertainty on $\alpha$ is much larger in the first bins. In the second panel, we see that the highlighted templates underestimate the uncertainty while the last panel represents the agreement of the various measurements computed in sigmas as $\left|\alpha-\alpha_\mathrm{ref}\right|\left(\sigma^2+\sigma_\mathrm{ref}^2\right)^{-1/2}$.}
    \label{fig:templates_comparison}
\end{figure*}
\begin{figure}
    \centering
    \resizebox{\hsize}{!}{\includegraphics{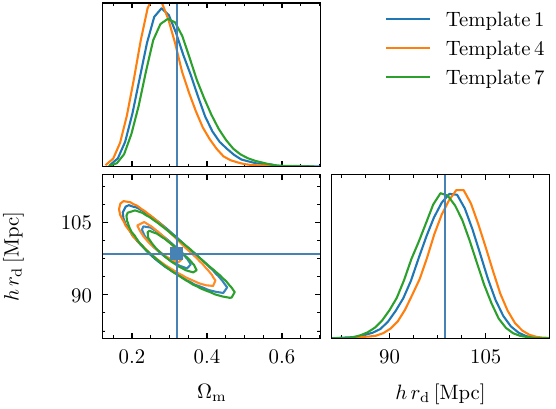}}
    \caption{Comparison of templates 1, 4, and 7 for BAO fitting at the level of cosmological parameters. Templates 7 and 8 show contours that are the least biased, whereas template 4 exhibits the largest shifts with $0.63\,\sigma$ and $0.45\,\sigma$ on \Omm\, and $h\,r_\mathrm{d}$. Other templates have similarly-biased contours.}
    \label{fig:templates_comparison_cosmo}
\end{figure}

The \,polynomial \,correction \,applied \,in \,the \,template \,is \,defined arbitrarily and different choices can be found in the literature. It is important to check whether the polynomial includes enough orders to absorb the shape of $w(\theta)$ and eventual non-linearities of the galaxy bias. In this context, we explored ten different combinations of orders leading to ten templates using the same scale cuts and fiducial cosmology as for the analysis of the individual bins. The templates considered in this analysis are
\vspace{3pt}
\begin{itemize}
\item template 1: $B \, w_{\mathrm{fid}}(\alpha \theta) + A_0 + A_1\,\theta^{-1} + A_2\,\theta^{-2}$,

\item template 2: $B \, w_{\mathrm{fid}}(\alpha \theta) + A_0 + A_1 \theta + A_2\,\theta^{-1}$,

\item template 3: $B \, w_{\mathrm{fid}}(\alpha \theta) + A_0 + A_1 \theta + A_2\,\theta^{-2}$,

\item template 4: $B \, w_{\mathrm{fid}}(\alpha \theta) + A_0 + A_1 \theta + A_2 \theta^2$,

\item template 5: $B \, w_{\mathrm{fid}}(\alpha \theta) + A_0 +  A_1\,\theta^{-1} + A_2 \theta^2$,

\item template 6: $B \, w_{\mathrm{fid}}(\alpha \theta) + A_0 + A_1\,\theta^{-1}$,

\item template 7: $B \, w_{\mathrm{fid}}(\alpha \theta) + A_0 + A_1\,\theta^{-1} + A_2\,\theta^{-2} + A_3\,\theta^{-3}$,

\item template 8: $B \, w_{\mathrm{fid}}(\alpha \theta) + A_0 +  A_1\,\theta^{-1} + A_2\,\theta^{-2} + A_3\,\theta^{-3} + A_4\,\theta^{-4}$,

\item template 9: $B \, w_{\mathrm{fid}}(\alpha \theta) + A_0 +  A_1\,\theta^{-1} + A_2\,\theta^{-2} + A_3 \theta $,

\item template 10: $B \, w_{\mathrm{fid}}(\alpha \theta) + A_0 +  A_1\,\theta^{-1} + A_2\,\theta^{-2} + A_3 \theta +  A_4 \theta^2$.
\end{itemize}
\vspace{3pt}
The comparison of the resulting $\alpha$ for each template with respect to the fiducial template 1 is shown in Fig.~\ref{fig:templates_comparison}. Constraints from templates 2, 4, 5, and 6 are highlighted because they are consistent with each other and significantly different from the reference at low redshift. These templates do not have a term in $1/\theta^2$. From this observation, it seems that this order is needed. We find that the $1\,\sigma$ uncertainty on $\alpha$ is systematically underestimated by 10\% with templates 2, 4, 5, and 6. This result is still observed when a noise-free synthetic data vector is used instead of the measured two-point correlation function. On the contrary, we see almost no variation in the results between the other templates with additional orders $\theta^{-3}, \theta^{-4}, \theta, \theta^2$, at the exception of templates 7 and, to a lesser extent, 8. These templates yield significantly larger uncertainties on $\alpha$ in bins 7 and 9, up to 46\%. Among these other templates, the agreement of the various measurements of $\alpha$, defined as $\left|\alpha-\alpha_\mathrm{ref}\right|\left(\sigma^2+\sigma_\mathrm{ref}^2\right)^{-1/2}$, remains within $0.15\,\sigma$ when varying the reference template. When cosmological constraints are inferred, as shown in Fig.~\ref{fig:templates_comparison_cosmo}, we find that templates 7 and 8 yield the least biased results. For template 8, the shifts are $0.07\,\sigma$ in \Omm\, and $0.23\,\sigma$ in $h\,r_\mathrm{d}$ , while for template 7 they are $0.12\,\sigma$ and $0.05\,\sigma$, respectively. The third least biased constraints are obtained with template 1, with shifts of $0.34\,\sigma$ and $0.18\,\sigma$, as mentioned above. Template 4 performs the worst, with biases of $0.63\,\sigma$ and $0.45\,\sigma$. Given the results at the level of $\alpha$ constraints, we nevertheless adopt template 1 for the main analysis. However, it would be interesting to investigate why template 7 leads to improved cosmological constraints.

\begin{table*}
\small
\centering
\setlength\tabcolsep{3pt}
\setcellgapes{1.5pt}\makegapedcells 
\caption{Values of $\alpha$ obtained with various templates using a fiducial cosmology that differs from the simulation cosmology ($h=0.73$ against $h=0.67$).}
\begin{tabular}{lccccccccccccccc}
\hline
\hline
\noalign{\vskip 3pt}
  Template  &  Bin 1  &  Bin 2  &  Bin 3  &  Bin 4  &  Bin 5  &  Bin 6  &  Bin 7  &  Bin 8  &  Bin 9  &  Bin 10  &  Bin 11  &  Bin 12  &  Bin 13  &  $\langle|1-\alpha|\rangle$  &  $\langle\sigma_{68}\rangle$  \\
\noalign{\vskip 1pt}
\hline
\noalign{\vskip 3pt}
     1      &  1.013  &  1.046  &  0.969  &  1.011  &  1.012  &  0.996  &  0.949  &  1.069  &  1.047  &  1.032  &  1.049   &  1.011   &  1.010   &              0.030               &             0.068             \\
     3      &  1.038  &  1.057  &  0.956  &  1.014  &  1.003  &  0.979  &  0.949  &  1.066  &  1.038  &  1.033  &  1.048   &  1.012   &  1.009   &              0.033               &             0.064             \\
     7      &  1.010  &  1.042  &  0.970  &  1.027  &  1.010  &  0.996  &  0.952  &  1.068  &  1.056  &  1.032  &  1.047   &  1.013   &  1.007   &              0.030               &             0.070             \\
     8      &  0.992  &  1.052  &  0.969  &  1.017  &  1.015  &  1.000  &  0.956  &  1.070  &  1.059  &  1.032  &  1.045   &  1.015   &  1.005   &              0.030               &             0.070             \\
     9      &  1.034  &  1.052  &  0.970  &  1.016  &  1.005  &  0.994  &  0.949  &  1.067  &  1.049  &  1.032  &  1.048   &  1.012   &  1.009   &              0.031               &             0.067             \\
     10     &  1.026  &  1.052  &  0.967  &  1.016  &  1.012  &  0.993  &  0.947  &  1.063  &  1.048  &  1.032  &  1.048   &  1.012   &  1.010   &              0.032               &             0.062             \\
\hline
\end{tabular}
\tablefoot{The templates used here are the reference template 1 and templates including higher polynomial orders. The columns $\langle|1-\alpha|\rangle$ and $\langle\sigma_{68}\rangle$ are, respectively, the bias and the uncertainty of $\alpha$ averaged over all redshift bins. We find that template 1 has the smallest $\langle\sigma_{68}\rangle$ among templates of minimum bias $\langle|1-\alpha|\rangle=0.03$.}
\label{tab:alpha_template_fiducial}
\end{table*}

 \begin{figure*}
    \centering
    \vspace{0.3cm}
    \includegraphics[width=0.9\textwidth]{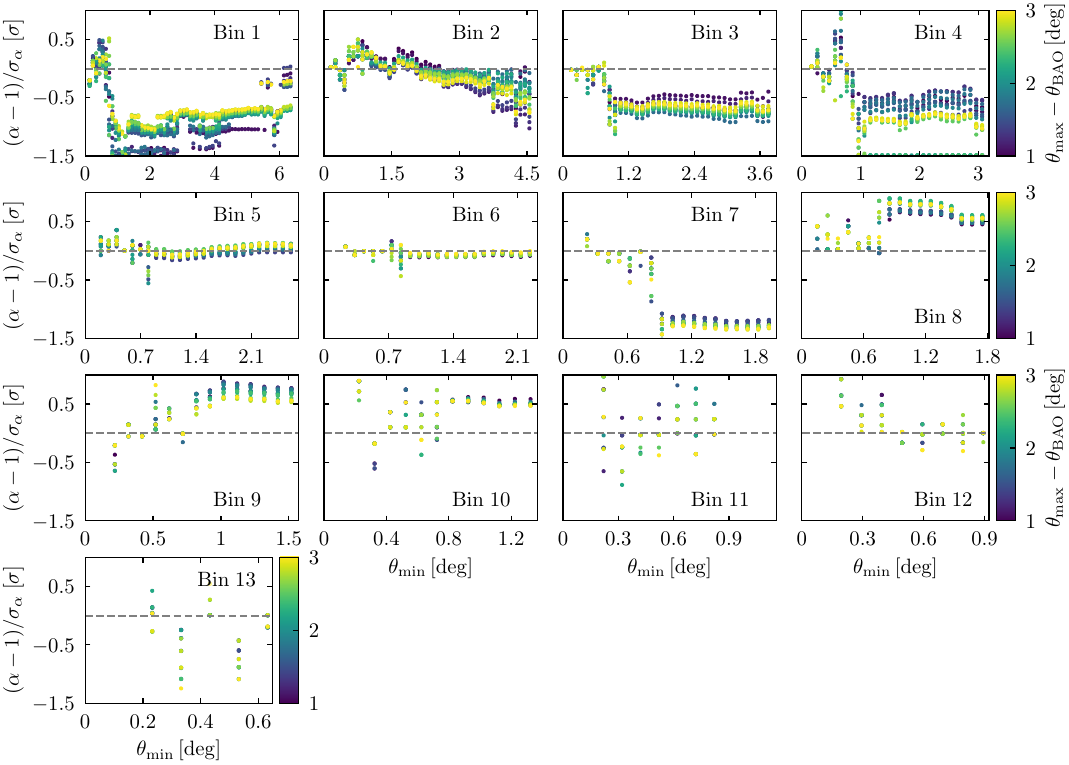}
    \caption{Bias $(\alpha-1)/\sigma_\alpha$ of $\alpha$ with respect to its uncertainty from Table~\ref{tab:alpha_i} as a function of $\thmin$ and $\thmax$ (colour) in the 13 redshift bins of the analysis. The dashed line corresponds to $\alpha-1=0$. In this analysis, $\alpha$ is obtained by best fit rather than MCMC. In almost all redshift bins, including small scales helps in recovering an unbiased estimate of $\alpha$.}
    \label{fig:fit_scale_cuts_study}
\end{figure*}

Another important aspect of the robustness of the template is to be able to handle shifts of the fiducial cosmology with respect to the true cosmology. As such, the previous comparison was repeated in a setup where the fiducial cosmology was altered to differ from the cosmology from the Flagship simulation, using $h=0.73$ instead of $h=0.67$. In this case, we find that the absolute shift of $\alpha$ with respect to the expected value of 1 averaged over all redshift bins $\langle|1-\alpha|\rangle$ for templates 1, 3, 7, 8, 9, 10 (we exclude templates without order $\theta^{-2}$) is, respectively, 0.030, 0.033, 0.030, 0.030, 0.031, and 0.032. Table~\ref{tab:alpha_template_fiducial} presents the results of this test, showing no preference for including higher orders in the template polynomial when it comes to robustness with respect to the choice of fiducial cosmology.

\subsection{Effect of the scale cuts} 
\label{sc:results_cuts}

The large redshift range $0.2 \leq z \leq 2.5$ included in this configuration space analysis entails that $\thbao$ varies between $1\degf6$ and $7^\circ$. In this situation, using a single scale cut for all bins or a redshift-dependent one is not obvious. We performed a first study of how $\alpha$ varies as a function of $\thmin$ and $\thmax$ with a fitting approach on a grid defined by $\thmin\in[0\degf2, \thbao-1^\circ]$ and $\thmax\in[\thbao+1^\circ, \thbao+3^\circ]$ by steps of $0\degf1$. The smallest data vector considered is $\thbao\pm1^\circ$ with 20 points. 

The results of this study are shown in Fig.~\ref{fig:fit_scale_cuts_study}. The bias $(\alpha-1)/\sigma_\alpha$ varies with $\thmin$, with a clear cut-off at $\theta=1^\circ$. Increasing $\thmin$ beyond this cut-off tends to bias $\alpha$, especially at low redshift, where the BAO signal is smeared by intra-bin variations of the BAO scale. While a pronounced evolution of $\alpha$ with $\thmax-\thbao$ is visible at low redshift when $\thmin\geq1^\circ$, no clear trend can be found when $\thmin\leq1^\circ$, where the bias is the smallest.
\vspace{3pt}

Following these observations, we then carried out an MCMC \vspace{1pt}study on a smaller grid defined as $\thmin\in[0\degf5,0\degf7]$ with the same \vspace{1pt}$0\degf1$ resolution and $\thmax\in[\thbao+1^\circ,\thbao+2\degf5]$ with steps of \vspace{1pt}$0\degf5$. On this grid, no significant variation in $\alpha$ was observed with \vspace{1pt}both $\thmin$ and $\thmax-\thbao$. The same observations were made when the analytical Gaussian covariance was used instead of the jackknife covariance. The final scale cut chosen was then $\thmin=0\degf6,\thmax=\thbao+2\degf5$, as it yielded the most robust results across all redshift bins. In the $\thmin, \thmax$ grid used for the MCMC, the $\alpha_i$ extracted in all bins are in agreement within $0.06\,\sigma$ on average and at most $0.25\,\sigma$.

\begin{figure}
    \vspace{4pt}
    \centering
    \includegraphics[width=0.45\textwidth]{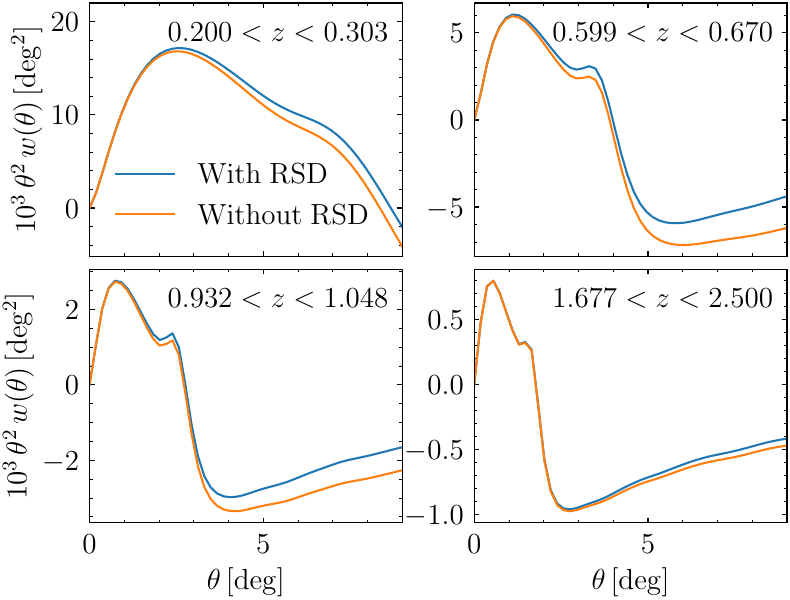}
    \caption{Theoretical two-point correlation function computed with and without RSDs using the redshift distributions of four redshift bins from the Flagship simulation shown in Fig.~\ref{fig:flagship_nz}.}
    \label{fig:rsd_effect_on_correlation}
\end{figure}

\subsection{Impact of RSDs}
\label{sc:results_rsd}
In this section, we checked whether including RSDs in the theoretical model described in Sect.~\ref{sc:theory_2pcf} affects the constraints on $\alpha$. The effect of RSDs over the two-point angular correlation function is illustrated in Fig.~\ref{fig:rsd_effect_on_correlation}. The correlation function amplitude is increased at large scales and low redshift. This is the Kaiser effect, explained by the inflow of galaxies towards overdensities, which is stronger at low redshift \citep{Kaiser_1987}.

However, the actual position of the BAO scale remains unchanged and we see that this is reflected in the constraints obtained from MCMC analysis. When RSDs are included, the template nuisance parameters compensate for the effect of RSDs with a significant decrease in $A_1$, while $A_0$ and $A_2$ are slightly increased. This trend holds in all redshift bins except the last one where the amplitude of the shifts becomes negligible. The relative difference $|\Delta\alpha|/\alpha$ averaged over all redshift bins is 0.25\%, the maximum being 0.5\%. As for the uncertainty of $\alpha$, its variations do not exceed 3.7\% and are on average 1.4\%. Thanks to the template-fitting approach, the analysis is robust with respect to RSDs. In any case, as explained in Sect.~\ref{sc:theory_2pcf}, RSDs are included in the analysis.

\subsection{Effect of the redshift binning scheme}
\label{sc:results_equidist_binning}

In this section, we show the effect of the choice of using equidistant redshift bins compared to the constraints obtained with equipopulated bins while keeping the same fiducial choice of template and scale cuts. This choice results in bins of width $\Delta z = 0.177$. For this analysis, the two-point correlation function and galaxy bias were measured in the same way as described in Sect.~\ref{sc:data_2pcf} for the equipopulated binning scheme. The rest of the analysis was carried out in an identical way. Using this binning scheme, we obtain equivalent constraints for redshift bins of similar $z_\mathrm{eff}$, as reported in Table~\ref{tab:alpha_i_equidist}. Note that the equidistant binning does not entail regularly spaced values of $z_\mathrm{eff}$. The effective redshifts are still computed with Eq.~\eqref{eq:effective_redshift}. By construction, we have more bins with high $z_\mathrm{eff}$ where the BAO signal is not smeared by the non-linear evolution of structures due to gravitation. Simultaneously, thinner redshift bins present a BAO signal that is less diluted by intra-bin variations in the BAO scale. However, the decrease in number density entails that the error bars of the two-point correlation function are larger at high redshift. In our case, this compromise yields more bins with high detection levels and tight constraints on $\alpha$. We also find several bins with low values of $\alpha$, two of them being more than $1\,\sigma$ away from $\alpha = 1$, at $z_\mathrm{eff}=1.993$ and 2.174. 

\begin{table}
\vspace{4pt}
\centering
\caption{Values of $\alpha$ extracted from MCMC in each equidistant redshift bin.}
\setcellgapes{0.5pt}\makegapedcells
\begin{tabular}{lcccc}
\hline
\hline
\noalign{\vskip 5pt}
 Bin &  $z_\mathrm{eff}$ & $\alpha$ & $D_\mathrm{A}/r_\mathrm{d}$ & $\Delta_\mathrm{det}$ ($\sigma$) \\
 \noalign{\vskip 1pt}
\hline
\noalign{\vskip 5pt}
 1 & 0.311 & $1.076^{+0.090}_{-0.145}$ & $7.165^{+0.599}_{-0.966}$ &       -- \\
 2 & 0.442 & $0.974^{+0.145}_{-0.114}$ & $8.085^{+1.204}_{-0.946}$ &      --   \\
 3 & 0.63  & $0.971^{+0.086}_{-0.089}$ & $9.645^{+0.854}_{-0.884}$ &      -- \\
 4 & 0.806 & $0.946^{+0.076}_{-0.063}$ & $10.331^{+0.830}_{-0.688}$ &      1.5 \\
 5 & 0.961 & $1.033^{+0.056}_{-0.060}$ & $11.871^{+0.644}_{-0.689}$ &      1.6 \\
 6 & 1.126 & $1.007^{+0.052}_{-0.049}$ & $11.963^{+0.618}_{-0.582}$ &      2.1   \\
 7 & 1.286 & $1.022^{+0.022}_{-0.022}$ & $12.365^{+0.266}_{-0.266}$ &      4.1 \\
 8 & 1.474 & $1.005^{+0.035}_{-0.032}$ & $12.277^{+0.428}_{-0.391}$ &      2.6 \\
 9 & 1.641 & $0.955^{+0.050}_{-0.042}$ & $11.678^{+0.611}_{-0.514}$ &      2.2 \\
 10 & 1.82  & $0.996^{+0.028}_{-0.028}$ & $12.124^{+0.341}_{-0.341}$ &       3.4 \\
 11 & 1.993 & $0.962^{+0.028}_{-0.025}$ & $11.613^{+0.338}_{-0.302}$ &       3.4 \\
 12 & 2.174 & $0.937^{+0.057}_{-0.045}$ & $11.179^{+0.680}_{-0.537}$ &       2.1   \\
 13 & 2.365 & $1.001^{+0.018}_{-0.018}$ & $11.768^{+0.212}_{-0.212}$ &      4.6  \\
\hline
\end{tabular}
\tablefoot{As there are more redshift bins at higher redshifts, the detection of the BAO signal is stronger than with equipopulated bins.}
\vspace{-6pt}
\label{tab:alpha_i_equidist}
\end{table}

Given these results, it seems that the equidistant scheme may be more suitable for the BAO analysis than the equipopulated scheme initially chosen to match the choice that maximises the dark energy figure of merit of the 3$\times$2-point analysis for \Euclid. In future \Euclid BAO photometric analyses, an optimisation of the photometric sample selection will be necessary.

\section{Conclusions}
\label{sc:Conclusions}
In this work, we have estimated our ability to constrain cosmological parameters with the photometric sample of \Euclid  through a BAO analysis of the Flagship galaxy mock catalogue, whose area is intermediate between the expected observed survey area at DR1 and DR2. We have measured the two-point correlation function in 13 redshift bins. We have extracted constraints on the BAO signal through the $\alpha$ parameter in each redshift bin but also using all bins jointly. The significance of the BAO detection was quantified to reach $\Delta_\mathrm{det}=10.3\,\sigma$ with the joint analysis, a three-fold improvement with respect to the latest results in photometric surveys \citepalias{Abbott_2024_DESY6}, with similar or better constraints in all redshift bins, while also covering a large range of redshifts, $0.2 \leq z \leq 2.5$.  This result shows how powerful photometric galaxy clustering can be with \Euclid. From these BAO constraints, we have derived constraints on different cosmological parameters: $\Omm=0.296^{+0.074}_{-0.059}$ and $h\,r_\mathrm{d}=99.35^{+4.29}_{-4.55}\,\mathrm{Mpc}$ in a flat $\Lambda$CDM cosmology.

We have also studied different analysis choices, showing that scale cuts can have a non-negligible effect on the value of the $\alpha$ parameter. Indeed, if \thmin$>$$1^\circ$, then we observe a bias, even for low-redshift bins where the BAO peak is at large scales. On the contrary, while omitting the order $\theta^{-2}$ in the template biases the resulting cosmological constraints, there is no need to include higher orders in the template functional form used to fit the BAO feature. The template is robust to the choice of fiducial cosmology, with variations $|\Delta\alpha|/\alpha$ not exceeding 0.0031 when averaged over all redshifts, well below the uncertainties of $\alpha$ given in Table~\ref{tab:alpha_i}. The template is efficient when it comes to absorbing effects that affect the amplitude of the two-point angular correlation function such as RSDs: including or removing them from the model, the constraints on $\alpha$ remain unchanged at all redshifts. An important point that was seen in both joint and individual redshift bin analyses is the robustness with respect to the redshift bins included in the analysis. We have also checked that there is no significant bias when a synthetic data vector is used. However, when we consider the data vector measured on the Flagship simulation, we have observed a shift of the $\alpha$ parameter and an increase in its uncertainty when redshift bin 11 is excluded. This effect is seen in both the joint analysis and the cosmological parameters inferred from $\alpha_{i}$, $i\in[3,13]$. This is explained by the fact that this bin has the largest BAO detection significance. This detailed study of the effect of each redshift bin will be mandatory in future works, given that the strength of the BAO signal and its effect on cosmological constraints change with redshift.

Results from Sect.~\ref{sc:results_equidist_binning} suggest that there is margin for improvement in the constraints obtained from a BAO analysis by optimising the redshift binning scheme. A galaxy sample selection taking into account both redshift and colour could be another possible point of improvement for these results. However, it is very promising to see that with a simulated area of 37\% of what is expected of \Euclid DR3, the constraints on $\alpha$ are already improved by a factor of three with respect to the current best constraints from a single photometric survey.
\\
\begin{acknowledgements}
\AckEC
This work has made use of CosmoHub, developed by PIC (maintained by IFAE and CIEMAT) in collaboration with ICE-CSIC. It received funding from the Spanish government (MCIN/AEI/10.13039/501100011033), the EU NextGeneration/PRTR (PRTR-C17.I1), and the Generalitat de Catalunya. Some of the results in this paper have been derived using the healpy and HEALPix packages. SC acknowledges support from the Italian Ministry of University and Research (\textsc{mur}), PRIN 2022 `EXSKALIBUR – Euclid-Cross-SKA: Likelihood Inference Building for Universe's Research', Grant No.\ 20222BBYB9, CUP D53D2300252 0006, and from the European Union -- Next Generation EU.
\end{acknowledgements}


\vspace{-0.39cm}
\begin{thebibliography}{62}
\expandafter\ifx\csname natexlab\endcsname\relax\def\natexlab#1{#1}\fi

\bibitem[{{Abbott} {et~al.}(2018){Abbott}, {Abdalla}, {Allam}, {et~al.}}]{DES_DR1}
{Abbott}, T.~M.~C., {Abdalla}, F.~B., {Allam}, S., {et~al.} 2018, \href{https://doi.org/10.3847/1538-4365/aae9f0}{\apjs, 239, 18}
\newpage
\bibitem[{Ade {et~al.}(2014)Ade, Aghanim, Arnaud, {et~al.}}]{Planck_2014}
Ade, P. A.~R., Aghanim, N., Arnaud, M., {et~al.} 2014, \href{https://doi.org/10.1051/0004-6361/201323003}{\aap, 566, A54}

\bibitem[{{Alonso} \& {Ferreira}(2015)}]{Alonso_2015}
{Alonso}, D. \& {Ferreira}, P.~G. 2015, \href{https://doi.org/10.1103/PhysRevD.92.063525}{\prd, 92, 063525}

\bibitem[{{Behroozi} {et~al.}(2013){Behroozi}, {Wechsler}, \& {Wu}}]{Behroozi_2013}
{Behroozi}, P.~S., {Wechsler}, R.~H., \& {Wu}, H.-Y. 2013, \href{https://doi.org/10.1088/0004-637X/762/2/109}{\apj, 762, 109}

\bibitem[{{Blanton} {et~al.}(2003){Blanton}, {Hogg}, {Bahcall}, {et~al.}}]{Blanton_2003}
{Blanton}, M.~R., {Hogg}, D.~W., {Bahcall}, N.~A., {et~al.} 2003, \href{https://doi.org/10.1086/375776}{\apj, 592, 819}

\bibitem[{{Blanton} {et~al.}(2005{\natexlab{a}}){Blanton}, {Lupton}, {Schlegel}, {Strauss}, {Brinkmann}, {Fukugita}, \& {Loveday}}]{Blanton_2005b}
{Blanton}, M.~R., {Lupton}, R.~H., {Schlegel}, D.~J., {et~al.} 2005{\natexlab{a}}, \href{https://doi.org/10.1086/431416}{\apj, 631, 208}

\bibitem[{{Blanton} {et~al.}(2005{\natexlab{b}}){Blanton}, {Schlegel}, {Strauss}, {et~al.}}]{Blanton_2005a}
{Blanton}, M.~R., {Schlegel}, D.~J., {Strauss}, M.~A., {et~al.} 2005{\natexlab{b}}, \href{https://doi.org/10.1086/429803}{\aj, 129, 2562}

\bibitem[{{Blas} {et~al.}(2011){Blas}, {Lesgourgues}, \& {Tram}}]{Diego_Blas_2011}
{Blas}, D., {Lesgourgues}, J., \& {Tram}, T. 2011, \href{https://doi.org/10.1088/1475-7516/2011/07/034}{JCAP, 07, 034}

\bibitem[{Carretero {et~al.}(2015)Carretero, Castander, Gaztañaga, Crocce, \& Fosalba}]{Carretero_2015}
Carretero, J., Castander, F.~J., Gaztañaga, E., Crocce, M., \& Fosalba, P. 2015, \href{https://doi.org/10.1093/mnras/stu2402}{\mnras, 447, 646}

\bibitem[{{Carretero} {et~al.}(2017){Carretero}, {Tallada}, {Casals}, {et~al.}}]{Carretero_2017}
{Carretero}, J., {Tallada}, P., {Casals}, J., {et~al.} 2017, in \href{https://doi.org/10.22323/1.314.0488}{Proceedings of the European Physical Society Conference on High Energy Physics. 5-12 July}, 488

\bibitem[{Challinor \& Lewis(2011)}]{Challinor_2011}
Challinor, A. \& Lewis, A. 2011, \href{https://doi.org/10.1103/PhysRevD.84.043516}{Phys. Rev. D, 84, 043516}

\bibitem[{Chisari {et~al.}(2019)Chisari, Alonso, Krause, {et~al.}}]{Chisari_2019}
Chisari, N.~E., Alonso, D., Krause, E., {et~al.} 2019, \href{https://doi.org/10.3847/1538-4365/ab1658}{\apjs, 242, 2}

\bibitem[{Cole {et~al.}(2005)Cole, Percival, Peacock, {et~al.}}]{Cole_2005}
Cole, S., Percival, W.~J., Peacock, J.~A., {et~al.} 2005, \href{https://doi.org/10.1111/j.1365-2966.2005.09318.x}{\mnras, 362, 505}

\bibitem[{Crocce {et~al.}(2011)Crocce, Cabré, \& Gaztañaga}]{Crocce_2011}
Crocce, M., Cabré, A., \& Gaztañaga, E. 2011, \href{https://doi.org/10.1111/j.1365-2966.2011.18393.x}{\mnras, 414, 329}

\bibitem[{{Dahlen} {et~al.}(2005){Dahlen}, {Mobasher}, {Somerville}, {et~al.}}]{Dahlen_2005}
{Dahlen}, T., {Mobasher}, B., {Somerville}, R.~S., {et~al.} 2005, \href{https://doi.org/10.1086/432027}{\apj, 631, 126}

\bibitem[{Dembinski {et~al.}(2020)Dembinski, Ongmongkolkul, Deil, {et~al.}}]{iminuit}
Dembinski, H., Ongmongkolkul, P., Deil, C., {et~al.} 2020, Zenodo, \href{https://doi.org/10.5281/zenodo.3949207}{\tt http://dx.\\doi.org/10.5281/zenodo.3949207}

\bibitem[{DES Collaboration(2024)}]{Abbott_2024_DESY6} DES Collaboration (Abbott, T.~M.~C., et~al.) 2024, \href{https://doi.org/10.1103/PhysRevD.110.063515}{\prd, 110, 063515}

\bibitem[{DESI Collaboration(2025a)}]{Adame_2025}
DESI Collaboration (Adame, A., et al.) 2025a, \href{https://doi.org/10.1088/1475-7516/2025/04/012}{JCAP, 04, 012}

\bibitem[{DESI Collaboration(2025b)}]{Abdul_Karim_2025}
DESI Collaboration (Abdul Karim, M., et al.) 2025b, \href{https://doi.org/10.1103/tr6y-kpc6}{\prd, 112, 083515}

\bibitem[{Eisenstein \& Hu(1998)}]{Eisenstein_1998}
Eisenstein, D.~J. \& Hu, W. 1998, \href{https://doi.org/10.1086/305424}{\apj, 496, 605}

\bibitem[{Eisenstein {et~al.}(2005)Eisenstein, Zehavi, Hogg, {et~al.}}]{Eisenstein_2005}
Eisenstein, D.~J., Zehavi, I., Hogg, D.~W., {et~al.} 2005, \href{https://doi.org/10.1086/466512}{\apj, 633, 560}

\bibitem[{{Euclid Collaboration: Blanchard} {et~al.}(2020)}]{Blanchard_2020}
Euclid Collaboration (Blanchard, A., et al.) 2020, \href{https://doi.org/10.1051/0004-6361/202038071}{\aap, 642, A191}

\bibitem[{{Euclid Collaboration: Pocino} {et~al.}(2021)}]{Pocino2021}
Euclid Collaboration (Pocino, A., et~al.) 2021, \href{https://doi.org/10.1051/0004-6361/202141061}{\aap, 655, A44}

\bibitem[{{Euclid Collaboration: Sciotti} {et~al.}(2024)}]{Sciotti_2024}
Euclid Collaboration (Sciotti, D., et~al.) 2024, \href{https://doi.org/10.1051/0004-6361/202348389}{\aap, 691, A318}

\bibitem[{{Euclid Collaboration: Castander} {et~al.}(2025)}]{Castander_2025}
Euclid Collaboration (Castander, F.~J., et al.) 2025, \href{https://doi.org/10.1051/0004-6361/202450853}{\aap, 697, A5}

\bibitem[{{Euclid Collaboration: Cropper} {et~al.}(2025)}]{Cropper_2025}
Euclid Collaboration (Cropper, M., et~al.) 2025, \href{https://doi.org/10.1051/0004-6361/202450996}{\aap, 697, A2}

\bibitem[{{Euclid Collaboration: Jahnke} {et~al.}(2025)}]{Jahnke_2025}
Euclid Collaboration (Jahnke, K., et~al.) 2025, \href{https://doi.org/10.1051/0004-6361/202450786}{\aap, 697, A3}

\bibitem[{{Euclid Collaboration: Mellier} {et~al.}(2025)}]{Mellier_2025}
Euclid Collaboration (Mellier, Y., et~al.) 2025, \href{https://doi.org/10.1051/0004-6361/202450810}{\aap, 697, A1}

\bibitem[{{Euclid Collaboration: Bertmann} {et~al.}(2026)}]{Bertmann_2026}
Euclid Collaboration (Bertmann, K.A., et~al.) 2026, \aap, submitted [arXiv:\href{https://doi.org/10.48550/arXiv.2604.00805}{\tt 2604.00805}]

\bibitem[{{Euclid Collaboration: Tucci} {et~al.}(2026)}]{Tucci_2025}
Euclid Collaboration (Tucci, M., et~al.) 2026, \href{https://doi.org/10.1051/0004-6361/202554588}{\aap, 711, A5}

\bibitem[{{Fang} {et~al.}(2020a){Fang}, {Krause}, {Eifler}, \& {MacCrann}}]{Fang_2020}
{Fang}, X., {Krause}, E., {Eifler}, T., \& {MacCrann}, N. 2020a, \href{https://doi.org/10.1088/1475-7516/2020/05/010}{JCAP, 05, 010}

\bibitem[{Fang {et~al.}(2020b)Fang, Eifler, \& Krause}]{Fang__2020}
Fang, X., Eifler, T., \& Krause, E. 2020b, \href{https://doi.org/10.1093/mnras/staa1726}{\mnras, 497, 2699}

\bibitem[{{Feroz} {et~al.}(2019){Feroz}, {Hobson}, {Cameron}, \& {Pettitt}}]{Feroz_2019}
{Feroz}, F., {Hobson}, M.~P., {Cameron}, E., \& {Pettitt}, A.~N. 2019, \href{https://doi.org/10.21105/astro.1306.2144}{Open J. Astrophys., 2, 10}

\bibitem[{Foreman-Mackey {et~al.}(2013)Foreman-Mackey, Hogg, Lang, \& Goodman}]{Foreman_Mackey_2013}
Foreman-Mackey, D., Hogg, D.~W., Lang, D., \& Goodman, J. 2013, \href{https://doi.org/10.1086/670067}{\pasp, 125, 306}

\bibitem[{{Gelman} \& {Rubin}(1992)}]{Gelman_1992}
{Gelman}, A. \& {Rubin}, D.~B. 1992, \href{https://doi.org/10.1214/ss/1177011136}{Stat. Sci., 7, 457}

\bibitem[{{G{\'o}rski} {et~al.}(2005){G{\'o}rski}, {Hivon}, {Banday}, {et~al.}}]{Gorski_2005}
{G{\'o}rski}, K.~M., {Hivon}, E., {Banday}, A.~J., {et~al.} 2005, \href{https://doi.org/10.1086/427976}{\apj, 622, 759}

\bibitem[{Gouyou~Beauchamps {et~al.}(2022)Gouyou~Beauchamps, Lacasa, Tutusaus, {et~al.}}]{Gouyou_Beauchamps_2022}
Gouyou~Beauchamps, S., Lacasa, F., Tutusaus, I., {et~al.} 2022, \href{https://doi.org/10.1051/0004-6361/202142052}{\aap, 659, A128}

\bibitem[{{Hartlap} {et~al.}(2007){Hartlap}, {Simon}, \& {Schneider}}]{Hartlap_2007}
{Hartlap}, J., {Simon}, P., \& {Schneider}, P. 2007, \href{https://doi.org/10.1051/0004-6361:20066170}{\aap, 464, 399}

\bibitem[{Hu \& Sugiyama(1996)}]{Hu_1996}
Hu, W. \& Sugiyama, N. 1996, \href{https://doi.org/10.1086/177989}{\apj, 471, 542}

\bibitem[{{Ivezi{\'c}} {et~al.}(2019){Ivezi{\'c}}, {Kahn}, {Tyson}, {et~al.}}]{Ivezic_2019}
{Ivezi{\'c}}, {\v{Z}}., {Kahn}, S.~M., {Tyson}, J.~A., {et~al.} 2019, \href{https://doi.org/10.3847/1538-4357/ab042c}{\apj, 873, 111}

\bibitem[{James \& Roos(1975)}]{James:1975dr}
James, F. \& Roos, M. 1975, \href{https://doi.org/10.1016/0010-4655(75)90039-9}{Comput. Phys. Commun., 10, 343}

\bibitem[{Jarvis {et~al.}(2004)Jarvis, Bernstein, \& Jain}]{Jarvis_2004}
Jarvis, M., Bernstein, G., \& Jain, B. 2004, \href{https://doi.org/10.1111/j.1365-2966.2004.07926.x}{\mnras, 352, 338}

\bibitem[{{Kaiser}(1987)}]{Kaiser_1987}
{Kaiser}, N. 1987, \href{https://doi.org/10.1093/mnras/227.1.1}{\mnras, 227, 1}

\bibitem[{{Kaiser}(1992)}]{Kaiser_1992}
{Kaiser}, N. 1992, \href{https://doi.org/10.1086/171151}{\apj, 388, 272}

\bibitem[{{Kirkby} {et~al.}(2013){Kirkby}, {Margala}, {Slosar}, {et~al.}}]{Kirkby_2013}
{Kirkby}, D., {Margala}, D., {Slosar}, A., {et~al.} 2013, \href{https://doi.org/10.1088/1475-7516/2013/03/024}{JCAP, 03, 024}

\bibitem[{{Krause} \& {Eifler}(2017)}]{Krause_2017}
{Krause}, E. \& {Eifler}, T. 2017, \href{https://doi.org/10.1093/mnras/stx1261}{\mnras, 470, 2100}

\bibitem[{Lacasa \& Grain(2019)}]{Lacasa_2019}
Lacasa, F. \& Grain, J. 2019, \href{https://doi.org/10.1051/0004-6361/201834343}{\aap, 624, A61}

\bibitem[{{Landy} \& {Szalay}(1993)}]{Landy_1993}
{Landy}, S.~D. \& {Szalay}, A.~S. 1993, \href{https://doi.org/10.1086/172900}{\apj, 412, 64}

\bibitem[{Lange(2023)}]{Lange_2023}
Lange, J.~U. 2023, \href{https://doi.org/10.1093/mnras/stad2441}{\mnras, 525, 3181}

\bibitem[{{Laureijs} {et~al.}(2011){Laureijs}, {Amiaux}, {Arduini}, {et~al.}}]{Laureijs_11}
{Laureijs}, R., {Amiaux}, J., {Arduini}, S., {et~al.} 2011, ESA/SRE(2011) 12, [arXiv:\href{https://doi.org/10.48550/arXiv.1110.3193}{\tt1110.3193}]

\bibitem[{Lepori {et~al.}(2022)Lepori, Tutusaus, Viglione, {et~al.}}]{Lepori_2022}
Lepori, F., Tutusaus, I., Viglione, C., {et~al.} 2022, \href{https://doi.org/10.1051/0004-6361/202142419}{\aap, 662, A93}

\bibitem[{{Mather} {et~al.}(1999){Mather}, {Fixsen}, {Shafer}, {et~al.}}]{Mather_1999}
{Mather}, J.~C., {Fixsen}, D.~J., {Shafer}, R.~A., {et~al.} 1999, \href{https://doi.org/10.1086/306805}{\apj, 512, 511}

\bibitem[{Mead {et~al.}(2021)Mead, Brieden, Tröster, \& Heymans}]{Mead_2021}
Mead, \,A.\,~J., \,Brieden, \,S., \,Tröster, \,T.,\, \& \,Heymans, \,C. 2021, \href{https://doi.org/10.1093/mnras/stab082}{\mnras, \,502, 1401}

\bibitem[{Percival {et~al.}(2001)Percival, Baugh, Bland-Hawthorn, {et~al.}}]{Percival_2001}
Percival, W.~J., Baugh, C.~M., Bland-Hawthorn, J., {et~al.} 2001, \href{https://doi.org/10.1046/j.1365-8711.2001.04827.x}{\mnras, 327, 1297}

\bibitem[{{Planck Collaboration VI}(2020){Planck Collaboration VI.}}]{Aghanim_2018}
Planck Collaboration VI. 2020, \href{https://doi.org/10.1051/0004-6361/201833910}{\aap, 641, A6}

\bibitem[{{Potter} {et~al.}(2017){Potter}, {Stadel}, \& {Teyssier}}]{Potter_2017}
{Potter}, D., {Stadel}, J., \& {Teyssier}, R. 2017, \href{https://doi.org/10.1186/s40668-017-0021-1}{CompAC, 4, 2}

\bibitem[{Tallada {et~al.}(2020)Tallada, Carretero, Casals, {et~al.}}]{TALLADA2020100391}
Tallada, P., Carretero, J., Casals, J., {et~al.} 2020, \href{https://doi.org/https://doi.org/10.1016/j.ascom.2020.100391}{Astron. Comput., 32, 100391}

\bibitem[{{Tanaka} {et~al.}(2018){Tanaka}, {Coupon}, {Hsieh}, {et~al.}}]{Tanaka_2018}
{Tanaka}, M., {Coupon}, J., {Hsieh}, B.-C., {et~al.} 2018, \href{https://doi.org/10.1093/pasj/psx077}{\pasj, 70, S9}

\bibitem[{{Troxel} {et~al.}(2018){Troxel}, {Krause}, {Chang}, {et~al.}}]{Troxel_2018}
{Troxel}, M.~A., {Krause}, E., {Chang}, C., {et~al.} 2018, \href{https://doi.org/10.1093/mnras/sty1889}{\mnras, 479, 4998}

\bibitem[{{Tutusaus} {et~al.}(2020){Tutusaus}, {Martinelli}, {Cardone}, {et~al.}}]{Tutusaus2020}
{Tutusaus}, I., {Martinelli}, M., {Cardone}, V.~F., {et~al.} 2020, \href{https://doi.org/10.1051/0004-6361/202038313}{\aap, 643, A70}

\bibitem[{{Zehavi} {et~al.}(2011){Zehavi}, {Zheng}, {Weinberg}, {et~al.}}]{Zehavi_2011}
{Zehavi}, I., {Zheng}, Z., {Weinberg}, D.~H., {et~al.} 2011, \href{https://doi.org/10.1088/0004-637X/736/1/59}{\apj, 736, 59}

\bibitem[{Zonca {et~al.}(2019)Zonca, Singer, Lenz, Reinecke, Rosset, Hivon, \& Gorski}]{Zonca2019}
Zonca, A., Singer, L., Lenz, D., {et~al.} 2019, \href{https://doi.org/10.21105/joss.01298}{J. Open Source Software, 4, 1298}
\end{thebibliography}
\pdfbookmark[1]{References}{references}

\pdfbookmark[1]{Authors and affiliations}{authors}

\begin{appendix}
\nolinenumbers

\section{Computation of \texorpdfstring{$r_\mathrm{d}$}{rdrag}}
\label{ap:rdrag}
In our analysis, we computed $r_\mathrm{d}$ following the fast approximations described by equations (2) to (6) in \citet{Eisenstein_1998}. In these equations, we updated the CMB temperature to its value $T_\mathrm{CMB}=2.725\,\mathrm{K}$ reported in the final FIRAS CMB results \citep{Mather_1999}. We report here the expressions used for the redshift $\zeq$ and wavenumber of the particle horizon $k_\mathrm{eq}$ at matter-radiation equality, where $\Theta_{2.7} = \frac{T_\mathrm{CMB}}{2.7\,\mathrm{K}}$, as well as the expression used for $z_\mathrm{drag}$:
\begin{align}
&\zeq = 2.5\expo{4}\,\omm\Theta_{2.7}^{-4}-1;\label{eq:z_eq_eisenstein}\\
&k_\mathrm{eq} = 0.0746\,\omm  \Theta_{2.7}^{-2};\label{eq:k_eq_eisenstein}\\
&z_\mathrm{drag,b1} = 0.313\, \omm^{-0.419} \left(1 + 0.607\,\omm^{0.674}\right);\label{eq:z_drag_b1_eisenstein}\\
&z_\mathrm{drag,b2} = 0.238\, \omm ^{0.223};\label{eq:z_drag_b2_eisenstein}\\
&z_\mathrm{drag} = 1345 \,\frac{\omm^{0.251}}{1 + 0.659\,\omm^{0.828}}  \left(1 + z_\mathrm{drag,b1}\omb^{z_\mathrm{drag,b2}}\right).\\
\label{eq:z_drag_eisenstein}
\end{align}
Then we compute the distance $r_\mathrm{d}$ as
\begin{align}
&R_\mathrm{drag} \equiv \frac{3\rho_\mathrm{b}(z_\mathrm{drag})}{4\rho_\mathrm{\gamma}(z_\mathrm{drag})} = 31.5\,\omb \Theta_{2.7}^{-4} \frac{1000}{1 + z_\mathrm{drag}};\label{eq:R_drag_eisenstein}\\
&R_\mathrm{eq}  \equiv \frac{3\rho_\mathrm{b}(\zeq)}{4\rho_\mathrm{\gamma}(\zeq)} = 31.5\,\omb \Theta_{2.7}^{-4} \frac{1000}{1 + \zeq};\label{eq:R_eq_eisenstein}\\
&r_\mathrm{d} = \frac{2}{3k_\mathrm{eq}}\sqrt{\frac{6}{R_\mathrm{eq}}} \ln\left(\frac{\sqrt{1+R_\mathrm{drag}}+\sqrt{R_\mathrm{drag}+R_\mathrm{eq}}}{1+\sqrt{R_\mathrm{eq}}}\right).\label{eq:r_drag_eisenstein}
\end{align}
The prefactor 1345 in Eq.~\eqref{eq:z_drag_eisenstein} was taken from Eq.~(E-2) in \citet{Hu_1996} rather than the 1291 prefactor used in \citet{Eisenstein_1998} as it provides significantly better agreement with the results from the Boltzmann codes \texttt{CAMB} and \texttt{CLASS} \citep{Diego_Blas_2011}. Varying $h$, $\Omb$, or $\Omm$ in [0.6, 0.8], [0.039, 0.059], and [0.17, 0.37], respectively, while fixing the other parameters to the fiducial values of Flagship, we find an average relative difference $\frac{r_\mathrm{d,\,HS}}{r_\mathrm{d,\,CAMB}}-1$ of $-0.04\%$, 0.01\%, and 0.03\% using 1345 as a prefactor against $\frac{r_\mathrm{d,\,EH}}{r_\mathrm{d,\,CAMB}}-1$ of 2.7\%, 2.7\%, and 2.6\% using 1291.

\section{Statistics of Flagship \texorpdfstring{\zph}{photo-z}}
\label{ap:zph_stats}
In this appendix, we present complementary statistics regarding the photometric redshifts of the sample described in Sect.~\ref{sc:Data}. In addition to the normalised median absolute deviation $\sigma_\mathrm{NMAD}$ introduced in Eq.~\eqref{eq:sigma_nmad}, we consider the mean scaled bias
\begin{equation}
\Delta z_\mathrm{scaled}=\left\langle\frac{\zph-z_\mathrm{true}}{1+z_\mathrm{true}}\right\rangle
\label{eq:bias}
\end{equation}
and the outlier fractions at 5, 10, and 15\% $\eta_{0.05}$, $\eta_{0.10}$, and $\eta_{0.15}$ defined as the fraction of galaxies for which the scaled bias is larger than 0.05, 0.10, and 0.15. These metrics are grouped in Table~\ref{tab:zph_stats_dr3_equipop} for the main sample used in the majority of this work and defined with 13 equipopulated redshift bins. Table~\ref{tab:zph_stats_dr3_equidist} lists the same results for the sample from Sect.~\ref{sc:results_equidist_binning} with 13 equidistant bins and Table~\ref{tab:zph_stats_dr1_equipop} describes the properties of the DR1-like sample used at the end of Sect.~\ref{sc:results_individual_bins}.

\begin{figure*}
\centering
\includegraphics[width=0.82\textwidth]{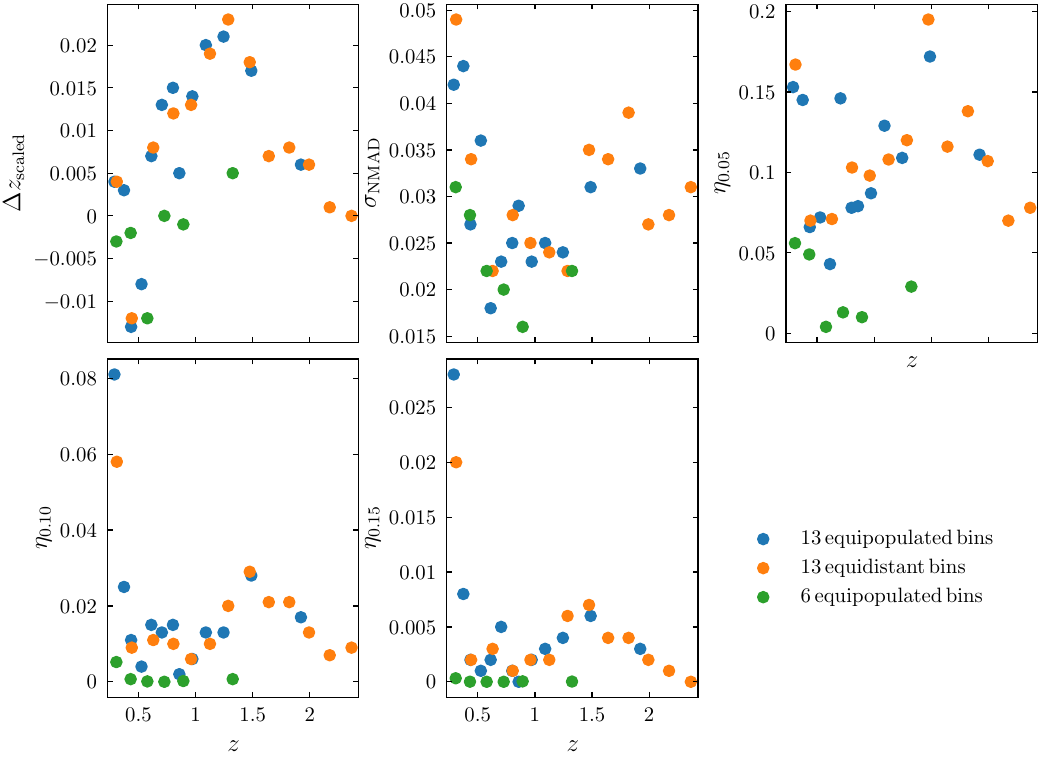}
\caption{Metrics of \zph\,as a function of the different bins effective redshift for the three galaxy samples considered in this work.}
\label{fig:photo_zs_stats_grouped}
\end{figure*}

These metrics are illustrated in Fig.~\ref{fig:photo_zs_stats_grouped} as a function of the effective redshift of the various bins used in the three galaxy samples. However, it is difficult to establish direct relationships between these metrics and the inferred values of $\alpha$.
\begin{table*}
\centering
\setlength\tabcolsep{5pt}
\caption{Statistics of the \zph\,of the sample defined with 13 equipopulated redshift bins used in the majority of this work.}
\begin{tabular}{lccccccccccccccc}
\hline
\hline
\noalign{\vskip 4pt}
 Metric                    &   Bin 1 &   Bin 2 &   Bin 3 &   Bin 4 &   Bin 5 &   Bin 6 &   Bin 7 &   Bin 8 &   Bin 9 &   Bin 10 &   Bin 11 &   Bin 12 &   Bin 13 \\
\hline
\noalign{\vskip 4pt}
 $\Delta z_\mathrm{scaled}$ &   0.004 &   0.003 &  $-$0.013 &  $-$0.008 &   0.007 &   0.013 &   0.015 &   0.005 &   0.014 &   0.02  &    0.021 &    0.017 &    0.006 \\
 $\sigma_\mathrm{NMAD}$    &   0.042 &   0.044 &   0.027 &   0.036 &   0.018 &   0.023 &   0.025 &   0.029 &   0.023 &   0.025 &    0.024 &    0.031 &    0.033 \\
 $\eta_{0.15}$             &   0.028 &   0.008 &   0.002 &   0.001 &   0.002 &   0.005 &   0.001 &   < 0.001     &   0.002 &   0.003 &    0.004 &    0.006 &    0.003 \\
 $\eta_{0.10}$             &   0.081 &   0.025 &   0.011 &   0.004 &   0.015 &   0.013 &   0.015 &   0.002 &   0.006 &   0.013 &    0.013 &    0.028 &    0.017 \\
 $\eta_{0.05}$             &   0.153 &   0.145 &   0.066 &   0.072 &   0.043 &   0.146 &   0.078 &   0.079 &   0.087 &   0.129 &    0.109 &    0.172 &    0.111 \\
\hline
\end{tabular}
\label{tab:zph_stats_dr3_equipop}
\end{table*}

\begin{table*}
\centering
\setlength\tabcolsep{5pt}
\caption{Statistics of the \zph\,of the sample defined with 13 equidistant redshift bins used in Sect.~\ref{sc:results_equidist_binning}.}
\begin{tabular}{lccccccccccccccc}
\hline
\hline
\noalign{\vskip 4pt}
 Metric                     &   Bin 1 &   Bin 2 &   Bin 3 &   Bin 4 &   Bin 5 &   Bin 6 &   Bin 7 &   Bin 8 &   Bin 9 &   Bin 10 &   Bin 11 &   Bin 12 &   Bin 13 \\
\hline
\noalign{\vskip 4pt}
 $\Delta z_\mathrm{scaled}$ &   0.004 &  $-$0.012 &   0.008 &   0.012 &   0.013 &   0.019 &   0.023 &   0.018 &   0.007 &    0.008 &    0.006 &    0.001 &    < 0.001     \\
 $\sigma_\mathrm{NMAD}$     &   0.049 &   0.034 &   0.022 &   0.028 &   0.025 &   0.024 &   0.022 &   0.035 &   0.034 &    0.039 &    0.027 &    0.028 &    0.031 \\
 $\eta_{0.15}$              &   0.02  &   0.002 &   0.003 &   0.001 &   0.002 &   0.002 &   0.006 &   0.007 &   0.004 &    0.004 &    0.002 &    0.001 &    < 0.001     \\
 $\eta_{0.10}$              &   0.058 &   0.009 &   0.011 &   0.01  &   0.006 &   0.01  &   0.02  &   0.029 &   0.021 &    0.021 &    0.013 &    0.007 &    0.009 \\
 $\eta_{0.05}$              &   0.167 &   0.07  &   0.071 &   0.103 &   0.098 &   0.108 &   0.12  &   0.195 &   0.116 &    0.138 &    0.107 &    0.07  &    0.078 \\
\hline
\end{tabular}
\label{tab:zph_stats_dr3_equidist}
\end{table*}

\begin{table*}
\centering
\setlength\tabcolsep{5pt}
\caption{Statistics of the \zph\,of the sample defined with 6 equipopulated redshift bins used at the end of Sect.~\ref{sc:results_individual_bins}.}
\begin{tabular}{lcccccc}
\hline
\hline
\noalign{\vskip 4pt}
 Metric                     &   Bin 1 &   Bin 2 &   Bin 3 &   Bin 4 &   Bin 5 &   Bin 6 \\
\hline
\noalign{\vskip 4pt}
 $\Delta z_\mathrm{scaled}$ & $-$0.003   & $-$0.002  & $-$0.012  &  < 0.001     & $-$0.001  &  0.005  \\
 $\sigma_\mathrm{NMAD}$     &  0.031   &  0.028  &  0.022  &   0.02  &  0.016  &  0.022  \\
 $\eta_{0.15}$              &  0.0003 &  < 0.0001  &  < 0.0001      &   < 0.0001     &  < 0.0001  &  < 0.0001  \\
 $\eta_{0.10}$              &  0.0052  &  0.0007 &  0.0001 &   < 0.0001     &  0.0002 &  0.0007 \\
 $\eta_{0.05}$              &  0.056   &  0.049  &  0.004  &   0.013 &  0.01   &  0.029  \\
\hline
\end{tabular}
\label{tab:zph_stats_dr1_equipop}

\end{table*}

\section{Profile likelihood comparisons}
\label{ap:proflikecomp}
\begin{figure}
\centering
\includegraphics[width=0.42\textwidth]{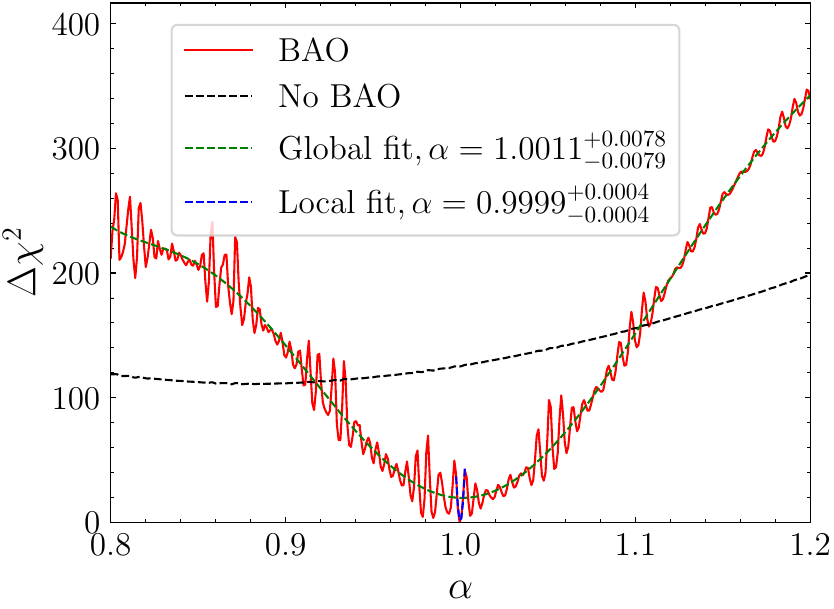}
\caption{Comparison of the profile likelihood from the joint analysis of all redshift bins.}
\label{fig:jointoscillations}
\end{figure}

In order to understand the oscillations observed in the posterior distribution of $\alpha$ in the individual bins and in the joint analysis, we used the profile likelihood approach since, in our case, MCMC poorly explores the $\alpha$ parameter space. The reason is illustrated in Fig.~\ref{fig:jointoscillations}, where the $\Delta\chi^2$ profile shown in red, with its associated polynomial fit in green, yields $\alpha=1.0011_{-0.0079}^{+0.0078}$. Restricting the range of the polynomial fit to the interval $\alpha\in[0.9965,1.0025]$, we were able to recreate the behaviour of the MCMC analysis, trapped in a local extremum, and obtain a value of $\alpha=0.9999 \pm 0.0004$ similar to the one obtained with MCMC.

\begin{figure}
\centering
\includegraphics[width=0.42\textwidth]{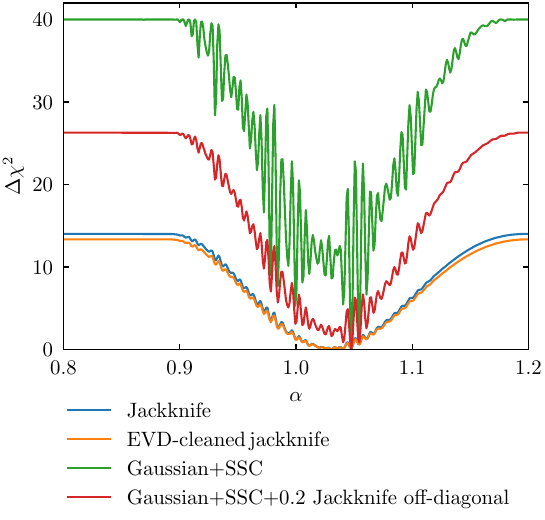}
\caption{Comparison of the influence of the covariance over the profile for bin 11.}
\label{fig:covariances}
\end{figure}

In the first analyses, we show the results obtained for bin 11 as it is the one with the largest oscillations in its posterior distribution, or alternatively, in the $\Delta\chi^2$ profile. However, most of the following tests were also repeated in the other bins with similar conclusions. The first sanity check performed was to compare three optimisation methods used to compute the $\Delta\chi^2$ profile over the range $\alpha\in[0.8,1.2]$: 
\begin{itemize}
\vspace{3pt}
\item the \texttt{mnprofile} function of the \texttt{iminuit} library, which \,directly returns the $\chi^2$ profile given the negative log-likelihood function and the grid of $\alpha$ values;
\item the \texttt{minimize} function of the \texttt{scipy} library, used to minimise the negative log-likelihood at each value of $\alpha$;
\item the same \texttt{scipy} minimisation approach but, following \citetalias{Abbott_2024_DESY6}, with analytical marginalisation over the nuisance parameters $A_i$.
\vspace{3pt}
\end{itemize}
The three methods yield rigorously the same result, as expected. We then compared the profiles obtained with various covariance matrices: the\,jackknife covariance used for the analysis of individual bins and a variant where the covariance is cleaned by removing eigenvalues smaller than $5\times10^{-10}$ (eigenvalue decomposition, EVD). This method allows for the inversion of the empirical covariance of the 13 redshift bins, enabling a joint analysis. We also considered the analytical Gaussian covariance with SSC used for the joint analysis and a variant where we added 20\% of the jackknife covariance off-diagonal terms. The choice of this last variant is motivated by the fact that off-diagonal terms are close to zero in the analytical covariance and may underestimate the covariance between data points. As shown in Fig.~\ref{fig:covariances}, the local minima are extremely deep with the analytical covariance. They are significantly shallower when a fraction of off-diagonal terms of the jackknife covariance is added and almost completely smoothed out with the jackknife covariances, which is consistent with their respective constraining power. However, the local minima are present with all covariances at the exact same $\alpha$ values in the profile so the covariance type is not the cause of the issue. Note that each $\Delta\chi^2$ profile was computed by removing its own minimum $\chi^2$ so that we clearly see each profile and the fact that oscillations are at the same positions. 

\begin{figure}
\centering
\includegraphics[width=0.42\textwidth]{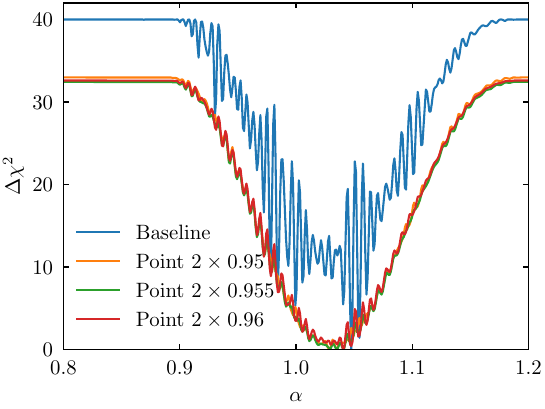}
\caption{Comparison of the influence of the second data point over the profile of bin 11 when it is shifted.}
\label{fig:point2shift}
\end{figure}

If the measurement from the Flagship simulation is replaced with a noiseless synthetic data vector, the profile becomes completely smooth, which suggests that the data is an important element in the explanation of this issue. To estimate the effect of each data point, we computed the profile while removing one point at a time. The most significant effect is observed when the second data point (at $\theta=0\degf87$) is ignored while removing other points yields variants of the profiles with the largest oscillations. In the other redshift bins, it is always one of the first points, at $\theta<1^\circ$, which induces the strongest oscillations and in most cases, the adjacent points also have a significant effect. Instead of removing completely the second data point in bin 11, we shifted it in Fig.~\ref{fig:point2shift} to verify if it is an outlier for the data despite the absence of visible outlier behaviour in Fig.~\ref{fig:flagship_2pcf}. We applied a factor towards larger or smaller values incrementally before finding a value for which the recovered profile is the smoothest. We find that this occurs for a value of the second data point which is 4.5\% smaller, equivalent to a 0.6$\sigma$ shift where $\sigma$ is the uncertainty of the data point. In this case, the median peak-to-peak amplitude of the $\chi^2$ oscillations decreases from 4.6 to 0.8. As a consequence, it seems that the data point is not an outlier from the point of view of the data. Instead, it is from the point of view of the likelihood surface that this point is an outlier, at a position that can favour different values of model parameters if shifted towards lower or higher values.

\begin{figure}
\centering
\includegraphics[width=0.42\textwidth]{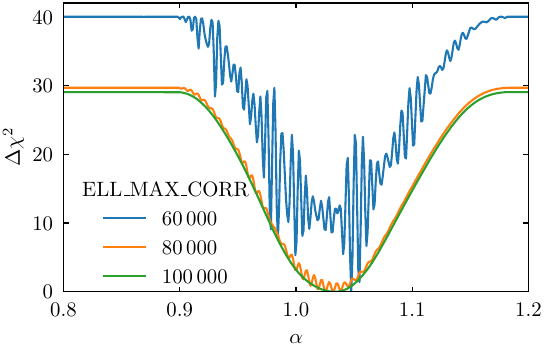}
\caption{Profile obtained for bin 11 when varying the Core Cosmology Library internal parameter $\mathrm{ELL\_MAX\_CORR}$.}
\label{fig:ellmaxcorr}
\end{figure}

We also investigated the template-fitting model used. We checked that using template 7 instead of 1 (adding a term in $1/\theta^3$) does not alleviate this issue. We also compare template 1 to the model used in the DES Y6 BAO analysis of \citetalias{Abbott_2024_DESY6}. This time, the associated profile is smooth, even with the analytical covariance. Further investigations revealed that the explanation of these oscillations is the Core Cosmology Library internal parameter $\mathrm{ELL\_MAX\_CORR}$. It controls the maximum $\ell$ of the spline used when computing $w(\theta)$ from angular power spectra with FFTLog. Its default value of 60\,000 had to be increased to 100\,000 to remove oscillations in redshift bin 11, as one can see from the $\Delta\chi^2$ profiles in Fig.~\ref{fig:ellmaxcorr}. To remove oscillations in all redshift bins, $\mathrm{ELL\_MAX\_CORR}$ was increased to 150\,000.

\begin{figure}
\centering
\includegraphics[width=0.42\textwidth]{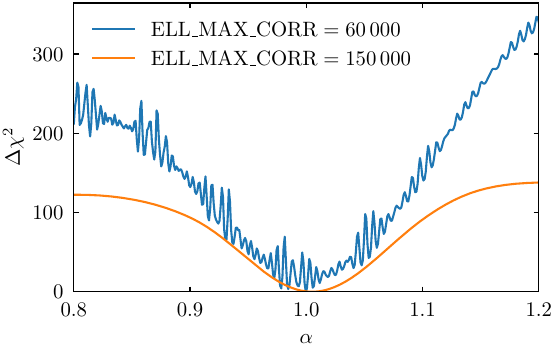}
\caption{Constraint on $\alpha$ from the joint analysis of all redshift bins using the default $\mathrm{ELL\_MAX\_CORR}=60\,000$ against 150\,000. The latter yields a smooth likelihood profile.}
\label{fig:joint}
\end{figure}

In the joint analysis, we find that using $\mathrm{ELL\_MAX\_CORR}=150\,000$ instead of 60\,000 considerably reduces the tension between the theoretical model and the measured data vector at $\theta<1^\circ$, as seen from the previous results. In Fig.~\ref{fig:joint}, we see that the profile is now smooth. The constraining power of the joint analysis obtained using this new parameter value is equivalent, yielding $\alpha=1.0055^{+0.0081}_{-0.0080}$. With MCMC, we find a 27\% tighter constraint $\alpha=1.0049\pm0.0059$. We note that with nested sampling, using the \texttt{nautilus} sampler \citep{Lange_2023,Feroz_2019}, we find $\alpha=1.0053_{-0.0079}^{+0.0078}$.

\end{appendix}
\end{document}